\documentclass[12pt, oneside]{book} 
\usepackage{amscd}
\usepackage{amsmath}
\usepackage{amsfonts}
\usepackage{amssymb}
\usepackage{amsthm}
\usepackage{extpfeil}
\usepackage{verbatim}
\usepackage[utf8]{inputenc}
\usepackage[margin=1in, a4paper]{geometry}               		
\usepackage[pdftex]{graphicx}	
\usepackage{standalone}  
\usepackage{setspace}
\usepackage{physics}
\usepackage{subcaption}
\usepackage[numbers]{natbib}
\usepackage{pdfpages}
\usepackage{bm}
\usepackage{wrapfig}				
\usepackage{longtable}
\usepackage{booktabs}
\usepackage{float}
\usepackage{afterpage}

\usepackage{bibentry} 
\usepackage[colorlinks=true, linkcolor=blue, urlcolor=blue, citecolor=blue, anchorcolor=blue]{hyperref}
\usepackage{siunitx}

\graphicspath{{images/}{{\subfix{images/}}}}
\usepackage{tikz}
\usepackage{tikz-3dplot}
\usetikzlibrary{shapes,calc,positioning, arrows.meta, quantikz2}
\tdplotsetmaincoords{70}{120}
\usepackage{siunitx}
\usepackage{adjustbox}
\usepackage{xcolor} 
\usepackage{blindtext}
\usepackage{rotating}
\usepackage{multicol,multirow}  
\usepackage{mathtools, nccmath}
\usepackage{caption}
\usepackage{subcaption}
\usepackage{subfiles} 

\usepackage{lipsum}			

\begin{document}

\begin{titlepage}
\begin{center}

\vspace*{1cm}

{\Large \textbf{A VQE simulation of the transverse field Ising model}\par} 

\vspace{0.8cm}
{\large Luke Bennett\par}
\vspace{0.3cm}

{\small
Department of Mathematical Sciences, University of Liverpool,\\
Liverpool L$69$ $7$ZL, United Kingdom\par}

\vspace{1.2cm}
\end{center}

\noindent\textbf{Abstract}

\vspace{0.2cm}

\noindent
One of the most promising applications of quantum computing is the ability to simulate quantum systems. A quantity of particular importance in condensed matter physics is the ground-state energy of a system. Here, we present an implementation of a hybrid quantum-classical algorithm known as the variational quantum eigensolver (VQE) to estimate the ground state energy of a prototypical quantum spin system. For this spin system, we use the transverse field Ising model (TFIM) which is known to exhibit a quantum phase transition (QPT) at a critical point. We apply the VQE to a $4$-spin TFIM and evaluate its performance using both a local noise model and IBM quantum hardware to estimate the ground-state energy and reproduce the QPT. The resulting ground-state energy estimates and QPT behaviour are compared with exact-diagonalisation benchmarks.

\vfill

\noindent
\textit{e-mail address:}
\href{mailto:luke.bennett@liverpool.ac.uk}{luke.bennett@liverpool.ac.uk}

\end{titlepage}

\frontmatter


\tableofcontents

\mainmatter

\chapter{Introduction}

Quantum computers are one of the most promising future technologies, standing to potentially revolutionise science, finance, cryptography and much more~\cite{Herman_2023, Gisin_2002}. Current devices are usually classed as noisy intermediate-scale quantum (NISQ) devices~\cite{Preskill_2018} and their usefulness for physics simulations is of particular interest. A class of algorithms
has emerged as a leading candidate for NISQ device physics simulations, known as variational quantum algorithms~\cite{Cerezo_VQA}. A specific variational algorithm of particular use to condensed matter physics is the variational quantum eigensolver (VQE)~\cite{PhysRevApplied.22.014055, Peruzzo_2014,Tilly_2022}.

A prototypical system for quantum simulation is the transverse field Ising model (TFIM). This simple quantum spin system exhibits interesting physical effects, namely quantum phase transitions (QPT)~\cite{sachdevQuantumPhaseTransitions2011, vojtaQuantumPhaseTransitions2003}. As a toy model, the TFIM is an excellent proof-of-concept for further quantum simulations~\cite{sharmaQuantumPhaseTransitions2026}\footnote{Published to the arXiv while this work was underway.}. Building on the methods used to simulate the TFIM, one can look to simulate more advanced spin systems~\cite{Bespalova_2025, mohtashimNeartermQuantumSimulation2025}. In addition to quantum simulations, the principles of spin systems have seen applications ranging from voter models to machine learning~\cite{article, 10660118}.

In this work we will first solve the TFIM classically using exact diagonalisation to calculate the ground state energy, as well as predict a QPT. We then apply the
VQE to a $4$-spin TFIM, beginning with a simple setup before performing a series of tests to reach a more optimised configuration. Using this we will generate estimates for the ground state energy using the VQE, as well as reproduce the QPT.

The work is organised as follows. We begin in~{Section}~\ref{sec:TFIM} with an introduction to the TFIM before proceeding with solving the model classically. We then, in~{Section}~\ref{sec:QC} introduce the fundamental concepts of quantum computing, before analysing the specificities of the VQE in~{Section}~\ref{sec:VQE}. Finally we present our simulation results in~{Section}~\ref{sec:results} and compare with those found classically before concluding in~{Section}~\ref{sec:conclusion} with our final remarks and recommendation for future work. All code and data used in this work are archived at Ref.~\cite{https://doi.org/10.5281/zenodo.20126040}. In addition, Ref.~\cite{Luke_git} provides an object-oriented implementation of the  original code designed for easier use and adaptation.

\chapter{Transverse Field Ising Model}
\newcommand{\gparam}{\ensuremath{g = \frac{h}{J}}}
\newcommand{\Hhat}{\ensuremath{\hat{H}}}

\label{sec:TFIM}
\section{Introduction} \label{sec:TFIM_intro}
The transverse field Ising model (TFIM) is a simple spin chain model used for the analysis of interacting quantum systems. The TFIM is categorised by its Hamiltonian, which is 
\begin{equation}\label{eq:hamiltonian_TFIM}
    \hat{H} = -J\sum_{i=1}^{N}\sigma_{i}^z \sigma_{i+1}^z
        -h\sum_{i=1}^{N}\sigma_{i}^x .
\end{equation}
This form of the Hamiltonian is for periodic boundary conditions, $\sigma_{N+1}^j = \sigma_{1}^j$. 
\begin{figure}[h]
    \centering
    \includegraphics[width=0.45\textwidth]{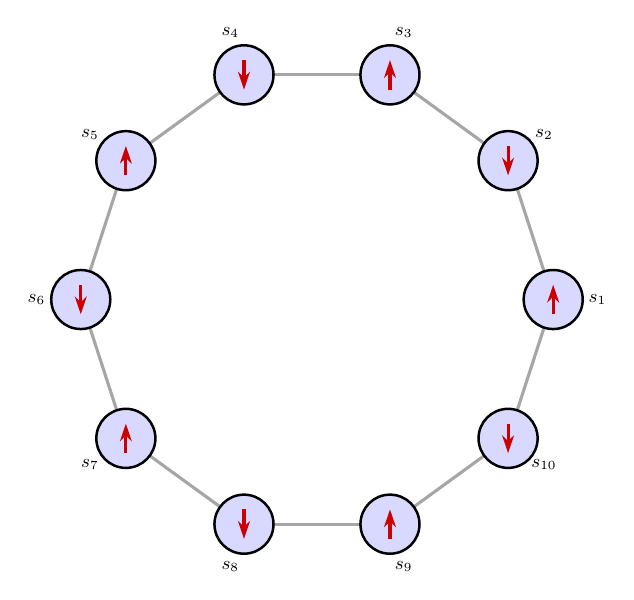}
    \caption{Spin chain with periodic boundary conditions, $N=10$ (generated using Claude Sonnet 4.6, \url{https://claude.ai}).}
    \label{fig:spinchain}
\end{figure}

This Hamiltonian consists of two opposing couplings, the spin-spin interaction $J$ and the transverse field interaction $h$. The first is responsible for the interactions between neighbouring spins within the chain. The constant $J$ is the magnetic interaction between the spins, which prefers their global ferromagnetic alignment when $J$ is positive and anti-ferromagnetic when $J$ is negative. The second term, $h$, is the  applied external transverse magnetic field, which disrupts the magnetic order \cite{sachdevQuantumPhaseTransitions2011}. For zero field, $h = 0$, the model
reduces to the well-known classical Ising model, with a ferromagnetic ground state \cite{vojtaQuantumPhaseTransitions2003}.

The Hamiltonian contains two Pauli spin operators, which form part of the larger Pauli basis:
\begin{align}
    \label{eq:Pauli_mat}
    \sigma^0 &=
    \begin{pmatrix}
    1 & 0 \\
    0 & 1
    \end{pmatrix}, &
    \sigma^x &=
    \begin{pmatrix}
    0 & 1 \\
    1 & 0
    \end{pmatrix}, &
    \sigma^y &=
    \begin{pmatrix}
    0 & -i \\
    i & 0
    \end{pmatrix}, &
    \sigma^z &=
    \begin{pmatrix}
    1 & 0 \\
    0 & -1
    \end{pmatrix}
\end{align}
The eigenvalues of $\sigma^{z}_i$ are $\pm 1$ which we associate with the two possible spin orientations, ``spin-up'' and ``spin-down'':
\begin{align}
    \ket{0} &= \ket{\uparrow}_i = \left(\begin{matrix}  1 \\ 0 \end{matrix}\right)_i, &
    \ket{1} &=  \ket{\downarrow}_i = \left(\begin{matrix}  0 \\ 1 \end{matrix}\right)_i.
\end{align}  

A key property of the Pauli operators within the TFIM Hamiltonian is that $\sigma^z$ and $\sigma^x$ do not commute when acting on the same site $i$. 
It is this non-commuting property that causes the quantum effects associated with a phase transition for the TFIM, as each operator is representative of the two competing terms within the Hamiltonian.

\section{Exact diagonalisation} \label{sec:ED}

One possible method for solving for the ground state energy of the TFIM is by diagonalizing the Hamiltonian. Although more advanced and efficient techniques exist, such as density-matrix renormalisation group methods~\cite{RevModPhys.77.259} for numerical studies and Jordan-Wigner transformations~\cite{mbengQuantumIsingChain2024} for analytic solutions, the systems considered in this work are limited to small spin chains $N = 2, 4$ and $8$. Consequentially, exact diagonalisation remains computationally feasible and provides a useful benchmark against which the VQE results can be compared. The reason  for this is that for spin systems there Hilbert space is exponentially large, dimension $2^N$ on a lattice with $N$ sites \cite{mbengQuantumIsingChain2024}, which clearly becomes computationally expensive for larger $N$. Once we have the Hamiltonian matrix it is a case of solving the time independent Schr\"odinger equation to find the ground state energy.

The first step we must follow in order to proceed exact diagonalisation is to represent the Hamiltonian~\eqref{eq:hamiltonian_TFIM} in matrix form. If we first consider the simple case of two spins in the chain, $N=2$, and setting $J=h=1$ we can expand the Hamiltonian:
\begin{equation}
    \hat{H} = - \sigma_{1}^z\sigma_{2}^z - (\sigma_{1}^x + \sigma_{2}^x).
\end{equation}
The lower indices in the above expression mean we apply the Pauli operator at that given site within the chain, i.e.~$\sigma_{1}^z$ means ``Apply the Pauli Z operator at the first site.'' In order to express this in matrix form we have to express each of the site acting Pauli operators as a single matrix equation. This is done is via the tensor product. Whenever we apply a Pauli operator to a given spin site, this has no effect on any of the other sites within the chain; mathematically, we can formulate this as applying the identity operator. For concreteness, let us focus on $\sigma_{1}^x$. If we now think of this as not just an operator acting on site $i=1$, but acting on the spin chain as a whole, then we can express this operator as $\sigma_{1}^x = \sigma^x \otimes I$. Similarly, we have $\sigma_{2}^x = I \otimes \sigma^x$. We can now express our Hamiltonian as a matrix equation:
\begin{equation}
\label{eq:H_intermediate}
    \hat{H} = - ( \sigma^z \otimes \sigma^z) - (\sigma^x \otimes I + I \otimes \sigma^x).
\end{equation}
Expanding~\eqref{eq:H_intermediate} gives,
\begin{equation}
\hat{H} = 
\left(
\begin{matrix}
-1 & -1 & -1 & 0\\
\\-1 & 1 & 0 & -1\\
\\-1 & 0 & 1 & -1\\
\\0 & -1 & -1 & -1
\end{matrix}
\right).
\end{equation}
This matrix is given in the $\sigma^z$ basis, such that the diagonal components correspond to the spin-interaction term within the Hamiltonian, whilst the off-diagonal components correspond to the field interaction term. This is one of the major benefits of using the $\sigma^z$ basis, as one can simply read off either the diagonal or off-diagonal components to see whether the effect of ``switching off'' one of the terms agrees with the classical analogue.

Now that we have the Hamiltonian in matrix form, we can proceed with the diagonalisation. For this, we make use of the Python library NumPy, in particular the function \texttt{eigh()}, which is a part of the built in linear algebra module and returns the eigenvalues and eigenvectors of a complex Hermitian (conjugate symmetric) or a real symmetric matrix \cite{harris2020array}. A useful feature of this function is that it automatically normalises the eigenvectors.

Following this procedure leads to the following values for the ground state energy when taking $h=J=1$.
\begin{table}[h]
    \centering
    \begin{tabular}{cc}
        \toprule
        $N$ & $E_0/J$ \\
        \midrule
        2 & -2.828 \\
        4 & -5.226 \\
        8 & -10.252 \\
        \bottomrule
    \end{tabular}
    \caption{Ground state energy $E_0$ for various system sizes $N$. The ratio $E_0/J$ is used to give dimensionless units.}
    \label{tab:ground_state_energy}
\end{table}

\section{Demonstrating a phase transition}\label{sec:ED_PT}

One of the most interesting properties of the TFIM is the models ability to demonstrate a quantum phase transition (QPT). Unlike classical phase transitions, which one can observe in a multitude of different physical systems, a QPT will occur at, or very close to,  absolute zero. Due to the absence of thermal energy, we conclude that all observed phenomena are driven by quantum effects. Before we discuss QPTs with application to the TFIM we shall first give a brief overview of phase transitions.

We typically put phase transitions into two different classes, first-order and continuous. First-order transitions are most likely what we have in our mind when we picture a phase transition. These are transitions for which two phases can exist simultaneously at the point of transition, e.g.~water turning into ice. Unlike first-order, continuous transitions are those for which the two phases cannot co-exist at the critical point. An important example of continuous phase transitions is the ferromagnetic transition of iron at $770^\circ C$ \cite{vojtaQuantumPhaseTransitions2003}, above which the material no longer exhibits strong magnetism, which is characteristic of ferromagnetism. In the context of the TFIM we will be concerned with continuous phase transitions.

When considering such continuous phase transition it will be useful to consider a so called order parameter. An order parameter is a mathematical quantity that takes a finite value when the system exists in the ordered phase, and vanishes in the disordered phase. When we say ordered vs disordered we are referring to the difference between the system being arranged in some regular/repeating pattern versus the contrary.

In the context of the TFIM the ordered phase is the (anti-)ferromagnetic phase (the type of ferromagnetism depends on the sign of the coupling constant $J$  however for simplicity we shall omit the `anti' for the rest of this discussion). It is clear that this is the ordered phase, as ferromagnetism is the result of parallel alignment of spins \cite{chikazumiPhysicsFerromagnetism1997}, which causes strong magnetic effects, such as those one would observe when placing a material such as iron near a permanent magnet.

In the Hamiltonian \eqref{eq:hamiltonian_TFIM} the first term, $-J\sum_{i=1}^{N}\sigma_{i}^z \sigma_{i+1}^z$, describes the ferromagnetic interaction along the spin $z$-axis. When we set the transverse field strength to $h=0$ in \eqref{eq:hamiltonian_TFIM}, the model reduces to the classical Ising model. In the $\sigma^z$ basis, this corresponds mathematically to a diagonal matrix with a ferromagnetic ground state. As we then sweep $h$, the system approaches a critical point due to the transverse field, which acts in the non-commuting $\sigma^x$ basis and causes the spin alignment to fluctuate. These are the quantum fluctuations mentioned above.

As we begin to sweep the transverse field $h$ whilst we keep the spin interaction $J$ fixed, we should at first notice no difference as for small $h$ the system should still be in the ferromagnetic phase. However, as $h$ increases it works to destroy this order by flipping the spins \cite{pfeutyOnedimensionalIsingModel1970} until the system reaches a critical point and the continuous phase transition has occurred, after this point the system will be in a disordered paramagnetic phase. 

\begin{figure}[h]
    \centering

    \begin{subfigure}{0.48\textwidth}
        \centering
        \includegraphics[width=\linewidth]{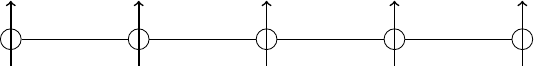}
        \caption{Ordered ferromagnetic phase}
        \label{fig:ferro}
    \end{subfigure}
    \hfill
    \begin{subfigure}{0.48\textwidth}
        \centering
        \includegraphics[width=\linewidth]{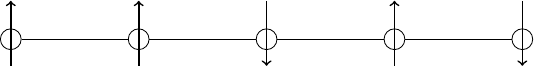}
        \caption{Disordered paramagnetic phase}
        \label{fig:para}
    \end{subfigure}

    \caption{Comparison of the two phases of the TFIM.}
    \label{fig:tfim_phases}
\end{figure}

We will now illustrate where the critical point for this continuous QPT will occur for the TFIM which we shall then use going forward in our analysis. Before we do this we shall note a couple of considerations; first the arguments below will be done within the infinite lattice limit, $N \to \infty$, which we will at times call the thermodynamic limit. Clearly no system that we can simulate analytically will have such limiting properties, however, we shall later show how the system's behaviour tends towards this limit for increasing $N$. Secondly, it is worth noting is that the conclusion we shall reach below can be reached through more advanced and rigorous methods, however we shall concentrate on perturbative methods.

We begin by defining a dimensionless parameter, 
\begin{equation}\label{eq:g_param}
g = \frac{h}{J},
\end{equation}
where $h$ and $J$ are the usual coupling constants ($J$ is again assumed positive to fix us working with ferromagnetism) as seen throughout this chapter, with both couplings having units of energy. The reason this parameter is useful is because if we were to scale the Hamiltonian, $\hat{H} \to \lambda \hat{H}$, the energies would change but the eigenstates and phase structure would not so $g$ is a control parameter which we can use to tune $\hat{H}$. Using this new parameter $g$ the Hamiltonian can be expressed as 
\begin{equation}\label{eq:Hamiltonian_g}
    \hat{H} = -J\sum_{i=1}^{N}\sigma_{i}^z \sigma_{i+1}^z
        -Jg\sum_{i=1}^{N}\sigma_{i}^x. 
\end{equation}
The ground state of $\hat{H}$ can depend only upon the value of the coupling $g$~\cite{sachdevQuantumPhaseTransitions2011}, as this represents the ratio of the two competing energy scales. We shall consider two different limits of $g$, first $g \gg 1$ which represents the transverse field term $h$ dominating, followed by $g \ll 1$ which likewise represents the spin interaction $J$ dominating.

\subsection[The large g limit]{The large $g$ limit} \label{sec:large_g_limit}

When considering the $g \gg 1$ limit we can form an approximation for the Hamiltonian~\eqref{eq:Hamiltonian_g},
\begin{equation}\label{eq:factorised_H_gg1}
    \hat{H} = -Jg \left[ \frac{1}{g} \sum_{i=1}^{N}\sigma_{i}^z \sigma_{i+1}^z
        + \sum_{i=1}^{N}\sigma_{i}^x\right] \approx -Jg \sum_{i=1}^{N}\sigma_{i}^x,
\end{equation}
to leading order in $1/g$. As the transverse field is the dominating term within this limit the spins will favour alignment with the spin $x$ axis, we shall denote the states of these alignments as $\ket{\pm}_i$, such states are the eigenstates of the $\sigma^{x}_i$ operator:
\begin{equation}
    \label{eq:paramag_eigenstates}
    \ket{\pm}_i = \frac{1}{\sqrt{2}}(\ket{\uparrow}_i \pm \ket{\downarrow}_i).
\end{equation}
The states~\eqref{eq:paramag_eigenstates} have corresponding eigenvalues $\pm 1$. Using this notation we can represent the ground state of the Hamiltonian in the disordered paramagnetic phase to leading order of $1/g$ as 
\begin{equation}
    \label{eq:paramag_gs}
    \ket{\Omega} = \prod_{i} \ket{+}_i .
\end{equation}

In order to understand how the spin system behaves in this limit, we need to understand how neighbouring spins are correlated with one another. We do this by using a correlation function. Correlation functions are used as a way to quantify how spins are related at different positions. In defining this correlation function we will be using $\sigma^{z}_i$ as these are the operators involved in the spin interaction. When in the state~\eqref{eq:paramag_gs} the values of $\sigma^{z}_i$ at different sites will be totally uncorrelated~\cite{sachdevQuantumPhaseTransitions2011} as a result of being in the disordered paramagnetic phase, the correlation function will thus be given as 
\begin{equation}
    \label{0_corr_func}
    \bra{\Omega}\sigma^{z}_i\sigma^{z}_j\ket{\Omega} = \delta_{ij}.
\end{equation}

 Perturbative corrections in $1/g$ will build in correlations in $\sigma^z$, representing spin interactions, that increase in range at each order~\cite{sachdevQuantumPhaseTransitions2011}. In order to quantify these correlations we introduce the correlation length, $\xi$, which is a mathematical quantity that quantifies the distance over which correlations exist; close to the critical point, the correlation length diverges \cite{vojtaQuantumPhaseTransitions2003}. Using this, we introduce a general result for the correlation function for order $n$~\cite{sachdevQuantumPhaseTransitions2011}:
 \begin{equation}
    \label{eq:paramag_corr_func}
    \bra{\Omega}\sigma_i^{z}\sigma_j^{z}\ket{\Omega} \sim e^{-\abs{x_i - x_j}/\xi},
\end{equation}
where $x_i$ is defined as the spatial coordinate for the lattice site $i$ (not to be confused with the $x$ associate with the transverse field axis).

The key detail in this correlation function is the exponential decay term, which implies that the correlations are expected to remain short ranged up to $n$ order perturbations. In saying short ranged we mean that there still isn't global alignment. As a result, neighbouring spins have only a small effect on the magnetic order which is quickly vanishing. Ultimately, this means that perturbative corrections in the current limit aren't enough to change the overall phase structure of the system, and we must instead consider the opposing limit, $g \ll 1$.

\subsection[The small g limit]{The small $g$ limit}

Now when considering the small $g$ limit we will find that the nature of the ground state is qualitatively different from the large-$g$ limit above \cite{sachdevQuantumPhaseTransitions2011}. We then use this to argue that the system must undergo a quantum phase transition between the two limiting cases at some critical point, $g = g_c$. Clearly, when considering this limit, the spin interaction term $J$ dominates, and the Hamiltonian has approximate form:
\begin{equation}
\label{eq:hamiltonian_g=0}
    \hat{H} \approx -J \sum_{i=1}^{N}\sigma^{z}_i\sigma^{z}_{i+1}.
\end{equation}

When $g=0$, which implies that $h=0$, the spins will be either all up or all down in the ground state, as there is no opposing transverse field to interrupt the interactions. We should note, there is still the possibility of small quantum fluctuations as a consequence of Heisenberg's uncertainty principle. However, these are much smaller in magnitude than the quantum fluctuations induced by the transverse field, and so can be neglected in the $h=0$ regime. As previously mentioned, the all up or all down orientations correspond to being in the ordered ferromagnetic phase. Thus, we can represent the two degenerate ground states in terms of the eigenstates of $\sigma^z$ as follows~\cite{sachdevQuantumPhaseTransitions2011}:
\begin{align}
    \label{eq:ferromag_gs}
    \ket{\uparrow} &= \prod_{i} \ket{\uparrow}_{i}, & 
    \ket{\downarrow} &= \prod_{i} \ket{\downarrow}_i .
\end{align}

By introducing a small but non-zero-$g$, such that $h \neq 0$, there will be the creation of a superposition of spin orientations. This superposition introduces some finite disorder into the system. The flipped states within this superposition, controlled by the finite perturbation in $g$, are generated through the associated transverse field operator, $\sigma^x$, which has the following action:
\begin{align}
    \label{eq:bit_flip}
    \sigma^x \ket{\uparrow}_i &= \ket{\downarrow}_i, &
    \sigma^x \ket{\downarrow}_i &= \ket{\uparrow}_i.
\end{align}
If we take $\ket{\Omega}$ to be the all-up ground state (we could have just as easily chosen all down as long as we are consistent with the orientations throughout) of site $i$, then we can see the effect of this perturbation on the ground state:
\begin{equation}
    \sigma^{x}_i \ket{\Omega} = \sigma^{x}_i \ket{\uparrow \uparrow \cdots \uparrow_i \cdots \uparrow} = \ket{\uparrow \uparrow \cdots \downarrow_i \cdots \uparrow}.
\end{equation}

Now, whilst this may feel like the obvious path towards the disordered paramagnetic phase, we must consider that we are working in the infinite lattice limit, $N \to \infty$, which means that the ground state degeneracy will survive at any finite order in a perturbation about $g=0$ \cite{sachdevQuantumPhaseTransitions2011}. This degeneracy of the ground states is due to the global $Z_2$ symmetry of the TFIM, which means that the Hamiltonian is invariant under the following global spin-flip transformation:
\begin{equation}
    \sigma^{z}_i \to - \sigma^{z}_i \quad \text{ for all sites }i.
\end{equation}

Since we cannot transform between the two ground states using any local operation, we say that such states are disconnected. When we say that we're transforming quantum states, physically, what is happening is that quantum tunnelling is taking place. In this case, since the two states are disconnected, no such tunnelling occurs. We should note that the absence of tunnelling is a distinct quality of being in the infinite lattice limit. In any practical application, there can always be some degree of tunnelling as a result of having a finite number of spins, although this will in general be exponentially suppressed.

When considering systems in the thermodynamic limit, such systems will always have to choose one or the other of the degenerate states as its ground state. As we have just shown, these two degenerate states are disconnected, and so in choosing a specific ground state we say that there has been spontaneous breaking of the $Z_2$ symmetry.

We can now follow a similar procedure to that of the large $g$ limit. We characterise the ground states by the considering the correlation function $\expval{\sigma^{x}_i \sigma^{z}_j}{\Omega}$, where again $\ket{\Omega}$ is either the all up or all down state. In order to find an expression for this correlation function we return to the Hamiltonian~\eqref{eq:hamiltonian_g=0}, which we know will have ground state $\ket{\Omega}$. To calculate the correlation function it will be useful to calculate the expectation value $\expval{\sigma^{z}_i}$, and in doing so we must first see how $\sigma^{z}_i$ acts on the possible ground states $\ket{\Omega}$:
\begin{align}
    \sigma^{z}_i \ket{\uparrow \uparrow \cdots \uparrow_i \cdots \uparrow} &= \ket{\uparrow \uparrow \cdots \uparrow_i \cdots \uparrow}, \\
    \sigma^{z}_i \ket{\downarrow \downarrow \cdots \downarrow_i \cdots \downarrow} &= -\ket{\downarrow \downarrow \cdots \downarrow_i \cdots \downarrow}.
\end{align}
Where the general Pauli bit operator can be calculated using the definition of the Pauli matrices~\eqref{eq:Pauli_mat}. If we then calculate the expectation  value of $\sigma^{z}_i$ with the ground state $\ket{\Omega}$ for $g=0$, we get
\begin{equation}
    \expval{\sigma^{z}_i} = \expval{\sigma^{z}_i}{\Omega} = \pm 1.
\end{equation}

If we were to then introduce some small quantum fluctuations into the system, i.e. a small factor of $g$, then by perturbation theory we find~\cite{sachdevQuantumPhaseTransitions2011}:
\begin{equation}
    \label{eq:spon_mag}
    \expval{\sigma^{z}_i} = \pm M_{0}(g), \quad \text{with} \quad M_{0}(0) = 1.
\end{equation}
What we have just derived is a perturbative expression for the spontaneous magnetisation $M_{0}(g)$, where higher order corrections in small $g$ reduce $M_0$ to a smaller, but non-zero, value~\cite{sachdevQuantumPhaseTransitions2011}. The spontaneous magnetisation is the magnetisation of a domain in zero applied magnetic field \cite{sucksmithSpontaneousMagnetizationTechniques1953}, so in our case this gives the inherent alignment of the spins within the ordered ferromagnetic phase. 

We can now characterise the ground states by the correlation function. Recalling that the correlation function is a measure of how neighbouring spins are related to one another. As the range becomes much larger, the correlation between such spins becomes weaker. As we approach the limit $\abs{x_i - x_j} \to \infty$, we can conclude that the spins must become independent. By continuing the small-$g$ perturbation, we are led to the following correlation function~\cite{sachdevQuantumPhaseTransitions2011}:
\begin{equation}
    \label{eq:ferromag_corr_func}
    \lim_{\abs{x_i - x_j} \to \infty} \expval{\sigma^{z}_i \sigma^{z}_j} = 
    \expval{\sigma^{z}_i} \expval{\sigma^{z}_j} = M_{0}^{2}.
\end{equation}

\subsection{The critical point}
Considering both of our derived correlation functions:
\begin{align}
    \expval{\sigma_i^{z}\sigma_j^{z}}{\Omega} &\sim e^{-\abs{x_i - x_j}/\xi}, &
    \expval{\sigma_i^{z}\sigma_j^{z}}{\Omega} &= M_0^2, 
\end{align}
we observe that there is no analytic method in which states that obey \eqref{eq:paramag_corr_func} can transform into \eqref{eq:ferromag_corr_func} as a function of the coupling $g$. As a result, we argue that there must be some critical point, $g=g_c$, which is necessarily non-analytic, at which the correlation function of the infinite lattice limit changes from long ranged (ferromagnetic) to short ranged (paramagnetic). This point corresponds to the QPT. As we have shown throughout this chapter, the two different phases occur at opposing limits of the control parameter, $g = h/J$, and so we conclude that the critical point of the QPT must occur when $h$ and $J$ have similar energy scales i.e. $h \sim J$. Thus, the control parameter we shall use going forward to demonstrate such a phase transition is 
\begin{equation}
    \label{eq:crit_point}
   \boxed{\frac{h}{J} \approx 1.} 
\end{equation}
Although we have shown this only as an approximate relation, the critical value of the control parameter can be derived exactly as $h/J = 1$ using renormalisation group methods.~\cite{sachdevQuantumPhaseTransitions2011}.

\begin{figure}[H]
    \centering
    \includegraphics[width=0.55\textwidth]{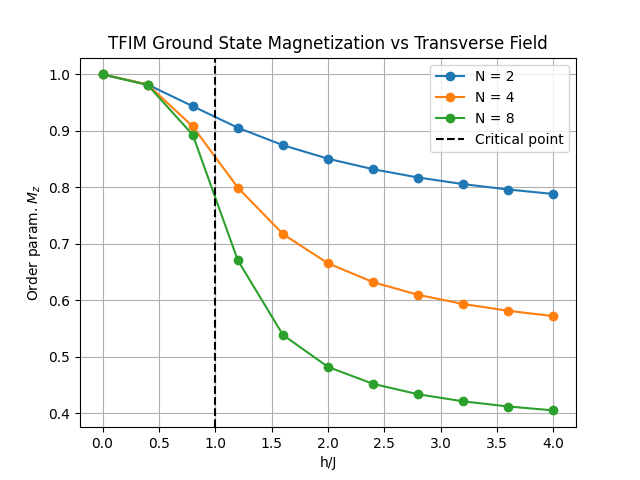}
    \caption{Order parameter plot for different numbers of spins $N$ with $J=1$ and $h$ swept from $0$ to $4$.}
    \label{fig:QPT_exact}
\end{figure}

Although the analysis above is for the infinite lattice limit, we are able to produce results that show the general behaviour of the critical point for the QPT with increasing accuracy for larger $N$. To do this, we must define the order parameter by considering the correlation function, which tends towards zero in the disordered paramagnetic phase and is non-zero in the ordered ferromagnetic phase. Such an order parameter, $M_z$, can be defined as:
\begin{equation}
    \label{eq:order_param}
    M_z = \frac{1}{N} \sqrt{\sum_{i,j} \expval{\sigma^{z}_i \sigma^{z}_j}} .
\end{equation}
In~\eqref{eq:order_param} we have normalised by a factor of $N$ to account for the number of spins within the chain. We demonstrate the general trend of the TFIM and its behaviour near the critical point in~{Fig.}~\ref{fig:QPT_exact}. Here, we observe that as $N$ increases, the curve gets closer to being discontinuous at the critical point. It will be the goal of this work to reproduce the curves in~{Fig.}~\ref{fig:QPT_exact} through the use of quantum variational methods.

Another useful quantity to measure is the time taken in calculating the order parameter for the full sweep of $h/J$ using the method of exact diagonalisation. 
\begin{table}[H]
    \centering
    \begin{tabular}{cc}
        \toprule
        $N$ & $t \, [s]$ \\
        \midrule
        2 & 0.017 \\
        4 & 0.081 \\
        8 & 1.050 \\
        \bottomrule
    \end{tabular}
    \caption{Time taken to calculate the magnetic order parameter using classical exact diagonalisation.}
    \label{tab:order_param_time}
\end{table}
From~{table}~\ref{tab:order_param_time}, we observe that the computation time increases rapidly with system size.This is consistent with the exponential scaling expected from classical exact diagonalisation and is one of the main drawbacks of using classical computers, especially for $N \gg 1$. This provides further motivation for the simulation of the TFIM through the use of a quantum algorithm.

\chapter{Quantum Computing}
\label{sec:QC}
\section{Introduction}
It has been known for some time that classical computers have a hard time simulating quantum physics beyond trivial systems. This is a consequence of the postulates of quantum mechanics, which require that physical quantities be represented as operators on a Hilbert space $\mathcal{H}$, the dimension of which scales exponentially with the number of particles. This problem led Richard Feynman, in a 1982 conference~\cite{feynmanSimulatingPhysicsComputers1982}, to propose using a quantum computer to simulate quantum systems, which classical computers cannot efficiently simulate. Over time, the original ideas of Feynman and others have evolved into the modern day pursuit of developing a universal quantum computer, a technological challenge requiring advances across physics, engineering and computer science.

In its most general form, quantum computation exploits key features of quantum mechanics, most notably superposition and entanglement. These phenomena underpin the potential advantages of quantum devices over their classical counterparts, as will be explored throughout this chapter.

\section{Foundations}\label{sec:QC_foundations}

To appreciate how a quantum computer works, it will first be useful for us to understand how classical computers work. From this, we can then outline how things differ as we move to the quantum case (as one also does when considering the difference between classical and quantum mechanics). At the most general level, a classical computer processes information using bits that can take values of either $0$ or $1$. These bits are physically realised through electronic circuits built from transistors, and modern computers contain billions of such components. An $N$ bit system can represent one of $2^N$ possible configurations, allowing classical computers to perform  tasks ranging from simple arithmetic to complex numerical calculations. However, for certain problems, such as advanced physics simulations or large integer factorisation, the computational resources required scale exponentially, placing such problems out of reach of classical computers.

Quantum computers offer a potential route to addressing some of these computational challenges. They operate based upon the superposition principle of quantum mechanics, which allows a quantum state $\ket{\Psi}$ to encode the information of a quantum system until one performs a measurement on $\ket{\Psi}$. At this point the system is projected onto a single state; this is the well-known collapse of the wave function. In the context of quantum computers this means that we replace a classical bit, which has a definite state of $0$ or $1$, with a quantum bit, the so-called qubit, which is a superposition of both $0$ and $1$. We can now promote the classical bit values to quantum states, which we shall refer to as the computational basis states, and represent each using Dirac notation: $\ket{0}$ and $\ket{1}$. As a result, we can form a mathematical expression for a single qubit:
\begin{equation}
\label{eq:single_qubit}
    \ket{\psi} = \alpha \ket{0} + \beta \ket{1}, \quad \text{where}, \quad 
    \alpha, \beta \in \mathbb{C}.
\end{equation}

 Only when an observer performs a measurement will a single output be given which will have an associated probability. For the single qubit case, if we obtain a measurement outcome of $0$, then this has an associated probability of $\abs{\alpha}^2$. It should then be clear that 
\begin{equation}
    \abs{\alpha}^2 + \abs{\beta}^2 = 1.
\end{equation}

From~\eqref{eq:single_qubit}, we can recognise a different interpretation of a qubit, that is, a two-dimensional vector within a complex Hilbert space $\mathcal{H}_2$, where the computational basis states form an orthogonal basis of $\mathcal{H}_2$. We can represent the computational basis states without loss of generality as:
\begin{align}
    \ket{0} &= \left(\begin{matrix}  1 \\ 0 \end{matrix}\right), &\quad
    \ket{1} &= \left(\begin{matrix}  0 \\ 1 \end{matrix}\right).
\end{align}
These expressions should be familiar from~{Sec.}~\ref{sec:TFIM_intro}, where the computational basis states were used to represent the spin orientation of a particle. Going forward, we shall think of the spins within the chain as qubits. Equipped with this new notion of qubits, one should also recall the states $\ket{\pm}$ from~{Sec.}~\ref{sec:large_g_limit}, which we now also consider as qubits within $\mathcal{H}_2$.

When considering the representation of a qubit in a two-dimensional complex Hilbert space, we can use the Bloch sphere to visualise the qubit's state. The Bloch sphere is a very usual tool in quantum computing, as it allows us to visualise both the state of a qubit and the action of certain operations on qubits. We shall see this explicitly later.
\begin{figure}[H]
    \centering
    \includegraphics[width=0.25\textwidth]{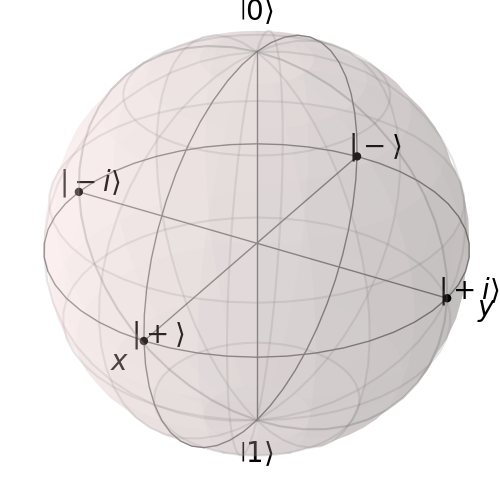}
    \caption{Bloch sphere made using QuTiP \cite{Johansson_2013}.}
    \label{fig:Bloch_sphere}
\end{figure}
In addition to the computational basis states and $\ket{\pm}$, we recognise two additional states on the Bloch sphere, $\ket{\pm i}$. These states are given as: 
\begin{equation}
\label{eq:i_states}
    \ket{\pm i} = \frac{1}{\sqrt{2}}\ket{0} \pm \frac{i}{\sqrt{2}}\ket{1}.
\end{equation}
The states \eqref{eq:i_states} have a similar structure to $\ket{\pm}$. The additional factor of $i$ introduces a relative phase of $\pi/2$, which rotates the state onto the $Y$-axis. The set of states $\{\ket{0}, \ket{1}, \ket{\pm}, \ket{\pm i}\}$ are referred to as ``pure states'', meaning that we have all of the information that characterises that state \cite{preskill2023quantum}. Unlike pure states, which exist on the surface of the Bloch sphere, any state within the Bloch sphere represents a mixed state, for which we only have partial information about the system.~\footnote{The definitions of pure and mixed states can be made mathematically concrete using the density matrix formalism of quantum mechanics.}

Whilst the Bloch sphere is an extremely useful visualisation tool, it must be kept in mind that this intuition is limited because there is no simple generalisation of the Bloch sphere known for multiple qubits~\cite[Chapter~1]{Nielsen_Chuang_2010}.

\section{Multiple qubits}\label{QC_mult_qubits}

Naturally, we would like to consider more than a single qubit. Classically, if we had, say, two bits, then their states could be described by the following state set: $\{00, 01, 10, 11 \}$, which contains all possible configurations of the system. Just like we did in the single qubit case, we can promote the classical states to a set of quantum computational basis states $\{\ket{00}, \ket{01}, \ket{10}, \ket{11} \}$. By leveraging the superposition principle, we can now represent the entire state vector of the two-qubit system as the following linear combination:
\begin{equation}
    \label{eq:2_qubit_LC}
    \ket{\psi} = \alpha_{00} \ket{00} + \alpha_{01} \ket{01} +
    \alpha_{10} \ket{10} + \alpha_{11} \ket{11}, \quad \text{where}
    \quad \alpha_{ij} \in \mathbb{C} .
\end{equation}
It is worth noting that writing expressions such as $\ket{00}$ are product states. Mathematically, these should be written as $\ket{0} \otimes \ket{0}$; however, we shall omit the tensor product symbol for simplicity. As the computational basis states can be perfectly distinguished they should be regarded as mutually orthogonal vectors in the Hilbert space of the composite system \cite{preskill2023quantum}, where this Hilbert space is denoted $\mathcal{H}_4 = \mathcal{H}_2 \otimes \mathcal{H}_2$ (the $4$ to indicate us being in some $4$-dimensional Hilbert space).

As in the single qubit case, only upon performing a measurement of the system will the state collapse to one of the computational basis states, with some probability $\abs{\alpha_j}^2$. This in turn leads to the following normalisation condition:
\begin{equation}
    \sum_{j} \abs{\alpha_j}^2 = 1.
\end{equation}
It is also possible to perform a measurement on a subset of qubits. For example, we could measure only the first qubit in a two-qubit state and observe an output of $1$, which would have an associated probability $\abs{\alpha_{10}}^2 + \abs{\alpha_{11}}^2$. The resulting post-measurement state is then:
\begin{equation}
    \ket{\psi'} = \frac{\alpha_{10} \ket{10} + \alpha_{11} \ket{11}}
    {\sqrt{\abs{\alpha_{10}}^2 + \abs{\alpha_{11}}^2}}.
\end{equation}
The denominator is just a re-normalisation factor to ensure that the probabilities correctly sum to unity. If we were then to measure the second qubit, we would observe a measurement outcome of either $0$ or $1$, with respective probabilities
\begin{align}
    P(0|1) &= {\frac{\abs{\alpha_{10}}^2}{\abs{\alpha_{10}}^2 + \abs{\alpha_{11}}^2}}, &
    P(1|1) &= \frac{\abs{\alpha_{11}}^2}{\abs{\alpha_{10}}^2 + \abs{\alpha_{11}}^2}. 
\end{align}

Some of the most important two-qubit states in quantum computing are the Bell states:
\begin{equation}
    \label{eq:Bell_states}
    \begin{aligned}
        \ket{\Phi^+} &= \frac{1}{\sqrt{2}}(\ket{00} + \ket{11}), \\
        \ket{\Phi^-} &= \frac{1}{\sqrt{2}}(\ket{00} - \ket{11}), \\
        \ket{\Psi^+} &= \frac{1}{\sqrt{2}}(\ket{01} + \ket{10}), \\
        \ket{\Psi^-} &= \frac{1}{\sqrt{2}}(\ket{01} - \ket{10}).
    \end{aligned}
\end{equation}
As we have four linearly independent states for this four-dimensional Hilbert space, the Bell states form a basis called the Bell basis. As a result, any state vector of a two-qubit system can be represented as a linear combination of Bell states:
\begin{equation}
    \label{eq:Bell_basis_LC}
    \ket{\psi} = \beta_{1} \ket{\Phi^+} + \beta_{2} \ket{\Phi^-} +
    \beta_{3} \ket{\Psi^+} + \beta_{4} \ket{\Psi^-}, \quad \text{where}
    \quad \beta_{j} \in \mathbb{C} .
\end{equation}
We say that the Bell states are maximally entangled quantum states, which means that they are perfectly correlated. We can demonstrate this by considering a single Bell state, $\ket{\Phi^+}$. Suppose we perform a measurement on the first qubit within this state and observe an output of $0$; this has probability $1/2$ and leaves the post measurement state
\begin{equation}
    \ket{\Phi'} = \ket{00}.
\end{equation}
Thus, we are guaranteed to observe an output of $0$ from the second qubit. Therefore, the states $\ket{\Phi^+}$ and $\ket{\Phi'}$ are said to be perfectly correlated. 

This statement has a profound impact, provided we have a Bell state, a measurement performed on one qubit immediately determines the outcome of a measurement on the other qubit. The implication of this phenomenon is that all of the information in a Bell state is stored in the correlations, not the individual qubits, this in turn allows quantum information to be distributed across a system rather than locally. This is a feature unique to quantum computers. This use of the entanglement principle  is the key ingredient in quantum teleportation and super-dense coding~\cite[Chapter~1]{Nielsen_Chuang_2010}, which are some of the most famous applications of quantum computing. Specific to the TFIM, the spins we consider in the chain are entangled through the spin-spin interaction term, thus the entanglement principle will be crucial for our later simulation.

We can generalise this discussion to a system of $n$-qubits. The Hilbert space for the $n$-qubit system is given as
\begin{equation}
    \label{eq:n_qubit_hilbert}
    \mathcal{H}_2^{\otimes n} = \mathcal{H}_2 \otimes \mathcal{H}_2 \otimes \dots 
    \otimes \mathcal{H}_2 \otimes \mathcal{H}_2.
\end{equation}
We can use \eqref{eq:n_qubit_hilbert} to calculate the dimension of the $n$-qubit system:
\begin{equation}
    \label{eq:n_qubit_dim}
    \dim(\mathcal{H}_2^{\otimes n}) = \prod_{i=1}^{n} \dim(\mathcal{H}_2) = 2^n.
\end{equation}
Thus, the possible states of this $n$-qubit system are vectors in this $2^n$-dimensional Hilbert space with complex coefficients~\cite{preskill2023quantum}. Such a space will have computational basis states of the form  $\ket{x_1 x_2 \dots x_n}$, where $ x \in \{0,1\}^n$. Here, $\{0,1\}^n$ denotes the set of all binary strings of length $n$. We now have all of the necessary ingredients to define a general $n$-qubit state vector:
\begin{equation}
    \label{eq:n_qubit_stateVec}
    \ket{\psi} = \sum_{x \in \{0,1\}^n} \!\! \alpha_x \ket{x}, \quad \alpha_x \in \mathbb{C}.
\end{equation}
Where the coefficients satisfy
\begin{equation}
    \sum_{x \in \{0,1\}^n} \abs{\alpha_x}^2 = 1.
\end{equation}

As a result of the Hilbert space having dimension $2^n$, the number of complex coefficients required to specify the full quantum state of a system quickly becomes unmanageable. For just an $n=500$ qubit system the number required is larger than the number of atoms within the universe~\cite[Chapter~1]{Nielsen_Chuang_2010}. Storing this amount of data on a classical computer requires exponential resources, which is what motivates quantum computation.

\section{Quantum gates and circuits}\label{sec:gates_cirucits}

We have now built up our understanding of qubits and the space in which they live, the Hilbert space $\mathcal{H}_2^{\otimes n}$. However, having qubits live in a Hilbert space where nothing happens isn't a very interesting situation. Naturally, the next question to ask is: how do we perform transformations on said qubits?

To answer this question, we must employ the concept of unitary transformations. Such transformations are one of the most important tools in quantum mechanics and play a fundamental role in quantum computation. Unitary transformations are defined to satisfy 
\begin{equation}
    \label{eq:unitary_trans}
    UU^{\dagger} = I.
\end{equation}
Operators satisfying \eqref{eq:unitary_trans} are called unitary operators, we also observe that \eqref{eq:unitary_trans} implies $U^{\dagger} = U^{-1}$ and so the ordering of the operators doesn't matter. To demonstrate the usefulness of unitary operators, suppose we have two states, $\ket{\psi_1}$ and $\ket{\psi_2}$, and we apply some unitary operator $U$ such that:
\begin{equation}
   \ket{\phi_1} = U\ket{\psi_1}, \quad \ket{\phi_2} = U\ket{\psi_2}. 
\end{equation}
Now, if we calculate the inner product $\bra{\phi_1 }\ket{\phi_2}$, we find
\begin{equation}
    \bra{\phi_1 }\ket{\phi_2} = \bra{U \psi_1} \ket{U \psi_2} = \bra{\psi_1}\ket{U^{\dagger}U\psi_2} = \bra{\psi_1}\ket{\psi_2}.
\end{equation}
We have just derived a famous result: the inner product is invariant under unitary transformations. A Hilbert space is a complete inner product space \cite{kreyszig1989introductory}, meaning that for a complex vector space to be called a Hilbert space it must have an associated well defined inner product. Hence, using unitary operators to transform qubits within $\mathcal{H}_2^{\otimes n}$ is the only way to ensure that we are always within the system's Hilbert space. 

When unitary transformations are applied to qubits they are called quantum gates and are the quantum computer’s counterpart of the elementary Boolean gates in the classical circuit model of computation \cite{preskill2023quantum}. We can also check if the inverse of a unitary quantum gate is itself, unitary. As $U^{-1} = U^{\dagger}$, 
\begin{equation}
    U^{-1}(U^{-1})^{\dagger} = U^{-1}(U^{\dagger})^{\dagger} = U^{-1}U = I.
\end{equation}
Thus, these quantum gates that we have now defined are always invertible. The same statement cannot be made for classical logic gates and so is a distinct feature of quantum computation.

\subsection{Single qubit gates}\label{subsec:single_qubit_gates}
In classical computing the only non-trivial logic gate, for single bits, is the NOT gate which maps either $0 \to 1$ or $1 \to 0$. Moving over to quantum computing, we would like to define an analogous gate, the quantum NOT gate. Recalling that the computational basis states can be represented as column vectors, we expect this NOT gate, henceforth simply referred to as the NOT gate rather than the quantum NOT gate, to be some $2 \times 2$ unitary matrix. This matrix is given by
\begin{equation}
    \label{eq:pauli_x}
    X = \begin{pmatrix}
                0 & 1 \\
                1 & 0
            \end{pmatrix}.
\end{equation}
We immediately recognise this as one of the Pauli matrices, specifically $\sigma^x$. Thus, this gate is also often called the Pauli-$X$ gate. We can verify that this gate does indeed perform the desired NOT operation:
\begin{equation}
    \begin{aligned}
       X\ket{0} &= \begin{pmatrix} 0 & 1 \\ 1 & 0 \end{pmatrix}
                    \begin{pmatrix} 1 \\ 0 \end{pmatrix}
                = \begin{pmatrix} 0 \\ 1 \end{pmatrix}
                = \ket{1}, \\[5pt]
        X\ket{1} &= \begin{pmatrix} 0 & 1 \\ 1 & 0 \end{pmatrix}
                    \begin{pmatrix} 0 \\ 1 \end{pmatrix}
                = \begin{pmatrix} 1 \\ 0 \end{pmatrix}
                = \ket{0}.
    \end{aligned}
\end{equation}
These operations also have a geometric interpretation. If we recall the Bloch sphere (see Figure~\ref{fig:Bloch_sphere}), one observes that the $X$-gate rotates a vector oriented in the direction of either of the computational basis states exactly $\pi$ radians. This has the effect of ``flipping'' the vector orientation; this gives us another name for the NOT gate, the bit-flip.

Additionally, we can consider how the bit-flip operates on the general qubit expression~\eqref{eq:single_qubit} by representing this as a vector in the computational basis:
\begin{equation}
    \label{eq:flip_qubit}
    X\ket{\psi} = \begin{pmatrix} 0 & 1 \\ 1 & 0 \end{pmatrix}
                  \begin{pmatrix} \alpha \\ \beta \end{pmatrix}
                = \begin{pmatrix} \beta \\ \alpha \end{pmatrix}.                   
\end{equation}
As expected, the bit-flip takes the state $\ket{\psi}$ to the corresponding state in which the role of $\ket{0}$ and $\ket{1}$ have been interchanged~\cite[Chapter~1]{Nielsen_Chuang_2010}. In general, the $X$ gate rotates a vector $\pi$ radians about the $X$-axis on the Bloch sphere.

The Pauli-$X$ is not the only Pauli matrix which is used as a quantum gate. We also use $\sigma^y$ and $\sigma^z$ as both are unitary matrices. These gates are given below:
\begin{align}
    \label{eq:pauli_YZ}
    Y &= \begin{pmatrix} 0 & -i \\ i & 0 \end{pmatrix}, &
    Z &= \begin{pmatrix} 1 & 0 \\ 0 & -1 \end{pmatrix}.           
\end{align}
We can apply each of these gates to the computational basis states, beginning with the $Z$ gate: 
\begin{equation}
    \label{eq:pauli_Z_action}
    \begin{aligned}
        Z\ket{0} &= \begin{pmatrix} 1 & 0 \\ 0 & -1 \end{pmatrix}
                    \begin{pmatrix} 1 \\ 0 \end{pmatrix}
                = \begin{pmatrix} 1 \\ 0 \end{pmatrix}
                = \ket{0}, \\[5pt]
        Z\ket{1} &= \begin{pmatrix} 1 & 0 \\ 0 & -1 \end{pmatrix}
                     \begin{pmatrix} 0 \\ 1 \end{pmatrix}
                = \begin{pmatrix} 0 \\ -1 \end{pmatrix}
                = -\ket{1}.
    \end{aligned}
\end{equation}
We notice that this gate has no effect on $\ket{0}$; however, when applied to $\ket{1}$, a phase factor of $-1 = e^{i\pi}$ is introduced, corresponding to a rotation of $\pi$ radians about the $Z$-axis on the Bloch sphere. As a result, the $Z$ gate is often referred to as a phase shift gate. When applied to the general expression~\eqref{eq:single_qubit}, the $Z$ gate performs the mapping
\begin{equation}
    \label{eq:Z_shift_qubit}
    \ket{\psi} = \alpha\ket{0} + \beta\ket{1} \xlongrightarrow{Z}
    \ket{\psi'} = \alpha\ket{0} + e^{i\pi}\beta\ket{1}.
\end{equation}
Whilst this clearly alters the relative phase between the two computational basis states, the overall normalisation, $\norm{\psi}^2$, remains constant preserving the superposition.

Considering the action of the $Y$ gate on the computational basis states, we obtain:
\begin{equation}
    \label{eq:pauli_YZ_action}
    \begin{aligned}
        Y\ket{0} &= \begin{pmatrix} 0 & -i \\ i & 0 \end{pmatrix}
                    \begin{pmatrix} 1 \\ 0 \end{pmatrix}
                = \begin{pmatrix} 0 \\ i \end{pmatrix}
                = i\ket{1}, \\[5pt]
        Y \ket{1} &= \begin{pmatrix} 0 & -i \\ i & 0 \end{pmatrix}
                     \begin{pmatrix} 0 \\ 1 \end{pmatrix}
                = \begin{pmatrix} -i \\ 0 \end{pmatrix}
                = -i\ket{0}.
    \end{aligned}
\end{equation}
The effect of this gate is both a bit flip combined with a phase shift, equivalent to a rotation of $\pi$ radians about the $Y$-axis. Applying the $Y$ gate to~\eqref{eq:single_qubit} performs the mapping
\begin{equation}
    \label{eq:Y_action_qubit}
    \ket{\psi} = \alpha\ket{0} + \beta\ket{1} \xlongrightarrow{Y}
    \ket{\psi'} = e^{-\frac{\pi}{2}i} \beta \ket{0} + 
    e^{+\frac{\pi}{2}i} \alpha \ket{1}.
\end{equation}

Sticking with the concept of Bloch sphere rotations, we can define more general gates: phase shift and rotation gates. The rotation gates allow one to specify an angle of rotation for a given qubit's state vector; successive rotation gates can then be applied to perform combinations of rotations. The rotation gates are given as:
\begin{equation}
    \label{eq:rotation_gates}
    \begin{gathered}
    R_x(\theta) = \begin{pmatrix}
        \cos\frac{\theta}{2} & -i\sin\frac{\theta}{2} \\
        -i\sin\frac{\theta}{2} & \cos\frac{\theta}{2}
        \end{pmatrix}, \\[10pt]
    R_y(\theta) = \begin{pmatrix}
        \cos\frac{\theta}{2} & -\sin\frac{\theta}{2} \\
        \sin\frac{\theta}{2} & \cos\frac{\theta}{2}
        \end{pmatrix}, \\[10pt]
    R_z(\theta) = \begin{pmatrix}
        \text{exp}\left(-i\frac{\theta}{2}\right) & 0 \\
        0 & \text{exp}\left(i\frac{\theta}{2}\right)
        \end{pmatrix}.
    \end{gathered}
\end{equation}
The expressions \eqref{eq:rotation_gates} will be very useful later on when using the variational quantum eigensolver. The phase shift gate is similar to the rotation gates, as one again specifies some angle. However, this angle is instead used to perform a phase shift on the $\ket{1}$ component. The phase shift gate is defined as:
\begin{equation}
    \label{eq:phase_shift}
    P(\phi) = \begin{pmatrix}
            1 & 0 \\
            0 & e^{i\phi}
        \end{pmatrix}
\end{equation}
Using \eqref{eq:phase_shift} we can derive the $Z$ gate for $\phi=\pi$. Additionally, we can define two common gates from the phase shift gate:
\begin{align}
    S &= P\left(\frac{\pi}{2}\right) = \begin{pmatrix}
        1 & 0 \\
        0 & i
    \end{pmatrix}, &
    T &= P\left(\frac{\pi}{4}\right) = \begin{pmatrix}
        1 & 0 \\
        0 & \frac{1+i}{\sqrt{2}}
    \end{pmatrix}.
\end{align}

Lastly, one of the most important gates used within quantum computing is the so-called Hadamard gate,\footnote{Named after the French mathematician Jacques Hadamard.}
\begin{equation}
    \label{eq:hadamard}
    H = \frac{1}{\sqrt{2}} \begin{pmatrix}
        1 & 1 \\
        1 & -1
    \end{pmatrix}.
\end{equation}
The Hadamard gate performs the following mappings:
\begin{align}
    \label{eq:hadamard_action}
        H\ket{0} &= \ket{+}, & H\ket{+} &= \ket{0}, \notag \\[5pt]
        H\ket{1} &= \ket{-}, & H\ket{-} &= \ket{1}.
\end{align}
From the Bloch sphere (see Figure~\ref{fig:Bloch_sphere}), we can identify the states used in~\eqref{eq:hadamard_action} as basis states: $\{\ket{0},\ket{1}\}$ for $Z$ and $\{\ket{+},\ket{-}\}$ for $X$. Thus, $H$ maps the $Z$ basis to the $X$ basis and vice versa. By applying a Hadamard gate successively, one can return to the original state,
\begin{equation}
    \ket{+} \stackrel{H}{\longrightarrow} \ket{0} 
    \stackrel{H}{\longrightarrow} \ket{+}.
\end{equation}
Hadamard gates are also used to create superpositions. Since it is 
often easiest to prepare a system in a computational basis state, 
applying a Hadamard gate then achieves superposition. On the Bloch 
sphere, the Hadamard operation corresponds to a rotation about the 
$Y$-axis by $\pi/{2}$ followed by a rotation about the $X$-axis 
by $\pi$~\cite[Chapter~1]{Nielsen_Chuang_2010}.

\subsection{Multiple qubit gates}\label{subsec:mult_qubit_gates}
Just as we built up our understanding of qubits from singular to multiple, we perform the same procedure when considering quantum gates. In classical computing there exists many different gates which act on multiple bits, common examples include AND, OR, XOR (exclusively or) and NAND (NOT-AND). We can introduce a similar class of quantum gates, which we shall refer to as control gates or control operations.

Suppose we have some arbitrary single qubit gate $U$, then a controlled-$U$ gate, often denoted $CU$, is a two-qubit operation that consists of a control and target qubit. If the control
qubit is set, then $U$ is applied to the target qubit; otherwise, the target qubit is left alone~\cite[Chapter~1]{Nielsen_Chuang_2010}. We can use this definition to construct a formula for a general controlled unitary operation. To do this, we use the following projectors: 
\begin{align}
    P_{0} &= \ket{0}\bra{0}, & P_{1} &= \ket{1}\bra{1}.
\end{align}
These projectors decompose the control-qubit Hilbert space into orthogonal subspaces, selecting the components corresponding to $\ket{0}$ and $\ket{1}$, respectively. Thus, when the control qubit is projected onto $\ket{0}$, the identity is applied to the target qubit, while projection onto ${\ket1}$ results in the application of $U$. This leads to the following formula:
\begin{equation}
    \label{eq:general_CU_opp}
    CU = \ket{0}\bra{0} \otimes I + \ket{1}\bra{1} \otimes U.
\end{equation}

If we consider the gate $U$ to be the single qubit NOT gate defined previously, then, using this definition of a controlled unitary operation, we can derive the action of the controlled-NOT ($CNOT/CX$) gate on the two-qubit computational basis states. Taking the first qubit to be the control qubit, we obtain:
\begin{align}
\label{eq_CNOT_map}
    \ket{00} &\to \ket{00}, & \ket{01} &\to \ket{01},
    &  \ket{10} &\to \ket{11}, &  \ket{11} &\to \ket{10}.
\end{align}
The actions of the CNOT gate that result in the mappings \eqref{eq_CNOT_map} can be represented as a single $4 \times4$ unitary matrix by applying \eqref{eq:general_CU_opp},
\begin{equation}
    \label{eq:CNOT_matrix}
    \text{CNOT} = \begin{pmatrix}
        1 & 0 & 0 & 0 \\
        0 & 1 & 0 & 0 \\
        0 & 0 & 0 & 1 \\
        0 & 0 & 1 & 0
    \end{pmatrix}.
\end{equation}
Additionally, the action of the CNOT gate can also be represented as $\ket{C T} \to \ket{C( T \oplus C)}$ ($C$ for control, $T$ for target), where $\oplus$ corresponds to addition modulo two. This representation is precisely what the classical XOR gate does. The CNOT gate, together with arbitrary single-qubit gates, gives the following universality result: any multi-qubit logic gate may be composed from CNOT and single qubit gates~\cite[Chapter~1]{Nielsen_Chuang_2010}.

The formula \eqref{eq:general_CU_opp} can be used to derive matrix representations for many different types of controlled unitary gates. Since \eqref{eq:general_CU_opp} is defined at the operator level, the formula can also be applied to an arbitrary qubit state. For example, if we consider a three-qubit state, one can construct a controlled-controlled-NOT (CCNOT) gate, commonly referred to as a Toffoli gate \cite{10.1007/3-540-10003-2_104}. Now, the third bit is a target bit that is flipped if both control bits are set to $1$; otherwise it is left unchanged~\cite[Chapter~1]{Nielsen_Chuang_2010}. Whilst we won't use the Toffoli gate in this work, it is fundamental to quantum computation as it can be used to simulate irreversible classical logic gates on a quantum computer~\cite{Bravyi_2016}.

Another useful multi-qubit gate is the swap gate, which has the action of swapping the positions of qubits in a state. In general, this performs the mapping
\begin{equation}
    \label{eq:gen_swap}
    \ket{ab} \to \ket{ba}.
\end{equation}
The swap gate can be represented in matrix form as:
\begin{equation}
    \label{eq:swap_mat}
    \text{SWAP} = \begin{pmatrix}
        1 & 0 & 0 & 0 \\
        0 & 0 & 1 & 0 \\
        0 & 1 & 0 & 0 \\
        0 & 0 & 0 & 1
    \end{pmatrix}.
\end{equation}

Below we summarise some of these key quantum operators along with their gate symbols, the symbols will be of particular importance as we construct quantum circuits. Note, for any of the control operations, the black circle represents the control qubit.

\begin{table}[h]
    \centering
    \begin{tabular}{ccc}
    \toprule
    Operator & Gate(s) & Matrix \\
    \midrule
    Pauli-X/NOT &
    $\begin{array}{c}
        \begin{quantikz} & \gate{X} & \end{quantikz} \\
        \begin{quantikz} & \targ{} & \end{quantikz}
    \end{array}$ &
    $\begin{pmatrix} 0 & 1 \\ 1 & 0 \end{pmatrix}$ \\[1em]
    Pauli-Y & \begin{quantikz} & \gate{Y} & \end{quantikz} &
    $\begin{pmatrix} 0 & -i \\ i & 0 \end{pmatrix}$ \\[1em]
    Pauli-Z & \begin{quantikz} & \gate{Z} & \end{quantikz} &
    $\begin{pmatrix} 1 & 0 \\ 0 & -1 \end{pmatrix}$ \\[1em]
    S & \begin{quantikz} & \gate{S} & \end{quantikz} &
    $\begin{pmatrix} 1 & 0 \\ 0 & i \end{pmatrix}$ \\[1em]
    T & \begin{quantikz} & \gate{T} & \end{quantikz} &
    $\begin{pmatrix} 1 & 0 \\ 0 & e^{i\pi/4} \end{pmatrix}$ \\[1em]
    Hadamard & \begin{quantikz} & \gate{H} & \end{quantikz} &
    $\frac{1}{\sqrt{2}} \begin{pmatrix} 1 & 1 \\ 1 & -1\end{pmatrix}$ \\
    CU & \begin{quantikz} & \ctrl{1} & \\ &\gate{U}& \end{quantikz} &
    $\begin{pmatrix} I & 0 \\ 0 & U \end{pmatrix}$ \\
    CNOT & \begin{quantikz} & \ctrl{1} & \\ &\targ{}& \end{quantikz} &
    $\begin{pmatrix}
        1 & 0 & 0 & 0 \\
        0 & 1 & 0 & 0 \\
        0 & 0 & 0 & 1 \\
        0 & 0 & 1 & 0
    \end{pmatrix}$ \\[2em]
    SWAP & \begin{quantikz} & \swap{1} & \\ &\targX{}& \end{quantikz} &
    $\begin{pmatrix}
        1 & 0 & 0 & 0 \\
        0 & 0 & 1 & 0 \\
        0 & 1 & 0 & 0 \\
        0 & 0 & 0 & 1
    \end{pmatrix}$ \\[2em]
    \bottomrule
    \end{tabular}
    \caption{Summary of some of the key quantum gates.}
    \label{tab:gates}
\end{table}

\subsection{Quantum circuits}\label{subsec:circuits}
Now that we have built up quantum gates, we can put these into practice and develop quantum circuits. In classical computers, circuits are used to represent information flow; the same is true for quantum computers. Quantum circuits are collections of quantum gates, which allow the flow of information; in this context, the information is encoded in qubits. We represent qubits via black lines, analogous to wires in classical circuits, with input qubits positioned on the left of the wire and output qubits at the right end of the wire. A trivial circuit is shown in~{Fig.}~\ref{fig:single_qubit_circuit}.

\begin{figure}[h]
    \centering
    \begin{quantikz}
    \lstick{$\ket{0}$} &&&&& \rstick{$\ket{0}$}
    \end{quantikz}
    \caption{Circuit representation of a single qubit.}
    \label{fig:single_qubit_circuit}
\end{figure}

\begin{figure}[h]
    \centering
    \begin{quantikz}
        \lstick{$\ket{\psi}$} && \meter{} & \setwiretype{c} &
    \end{quantikz}
    \caption{Measurement gate implemented in a simple circuit.}
    \label{fig:meas_gate_circuit}
\end{figure}
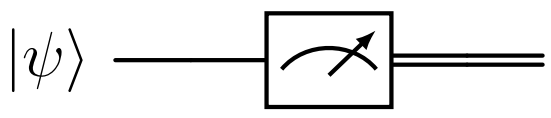

We can then begin to add gates from table \ref{tab:gates} to construct specific circuits. However, we must remember a fact of quantum mechanics: measurement outcomes always come with an associated probability. As a result, we can indicate at the end of a circuit what we anticipate the outcome to be, but until we perform a measurement, we have no way of knowing which outcome
will be observed. For example, suppose we construct a simple bit-flip circuit, which we illustrate in~{Fig}~\ref{fig:bit_flip_circuit}. Although we have indicated what we anticipate the outcome to be, we cannot say with certainty that this will indeed be the case. Only upon performing a measurement will we know whether  our circuit was successful. The circuit representation for a measurement is given in~{Fig.}~\ref{fig:meas_gate_circuit}, where the two horizontal lines represent that we are reading a classical output.

\begin{figure}[]
    \centering
    \begin{quantikz}
        \lstick{$\ket{0}$} && \gate{X} && \rstick{$\ket{1}$}
    \end{quantikz}
    \caption{Bit-flip circuit.}
    \label{fig:bit_flip_circuit}
\end{figure}
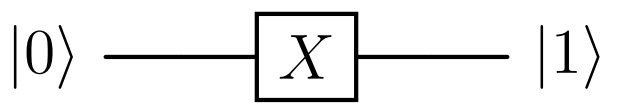

\begin{figure}[h]
    \centering
    \begin{quantikz}
        \lstick{$x$} & \gate{H} & \ctrl{1} && \\
        \lstick{$y$} && \targ{1} &&
    \end{quantikz}
    \caption{Circuit to prepare Bell state.}
    \label{fig:bell_state_circuit}
\end{figure}
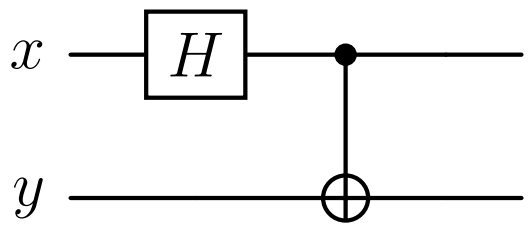

Building on these concepts, we can construct some more interesting circuits. The Bell states can be prepared via a specific quantum circuit. Suppose we have some quantum circuit that takes two qubits as input, $x$ and $y$. Together, these qubits form one of the standard two-qubit computational basis states: $\{\ket{00}, \ket{01}, \ket{10}, \ket{11}\}$. Then, if we construct a circuit consisting of a Hadamard gate and a CNOT gate, as shown in~{Fig.}~\ref{fig:bell_state_circuit}, we can use this to prepare one of the Bell states~\eqref{eq:Bell_states}.

Such a circuit can be decomposed into blocks that act on qubits at the same time. The first block consists of a Hadamard gate acting on the first qubit, whilst there is no action on the second qubit. As a result, we perform the tensor product:
\begin{equation}
    H \otimes I = \frac{1}{\sqrt{2}} \begin{pmatrix} 1 & 1 \\ 1 & -1\end{pmatrix}                   \otimes I
                = \frac{1}{\sqrt{2}} \begin{pmatrix} I & I \\ I & -I\end{pmatrix}
                = \frac{1}{\sqrt{2}} \begin{pmatrix} 1 & 0 & 1 & 0 \\ 0 & 1 & 0 & 1 \\ 1 & 0 & -1 & 0 \\ 0 & 1 & 0 & -1 \end{pmatrix}.
\end{equation}
The next block consists of a CNOT, whose matrix expression we already know (see table~\ref{tab:gates}). These can then be combined via matrix multiplication to give the matrix representation of the full circuit. To do this, we pre-multiply by the CNOT matrix and bring the scalar constant to the front:
\begin{equation}
    \frac{1}{\sqrt{2}} \begin{pmatrix}
        1 & 0 & 0 & 0 \\
        0 & 1 & 0 & 0 \\
        0 & 0 & 0 & 1 \\
        0 & 0 & 1 & 0
    \end{pmatrix}
    \begin{pmatrix} 1 & 0 & 1 & 0 \\ 0 & 1 & 0 & 1 \\ 1 & 0 & -1 & 0 \\ 0 & 1 & 0 & -1 \end{pmatrix} 
    = \frac{1}{\sqrt{2}}\begin{pmatrix} 1 & 0 & 1 & 0 \\ 0 & 1 & 0 & 1 \\ 0 & 1 & 0 & -1 \\ 1 & 0 & -1 & 0 \end{pmatrix}.
\end{equation}

Suppose that both input qubits are initialised in the state $\ket{0}$. The corresponding two-qubit input state is then
\begin{equation}
    \ket{00} = \ket{0} \otimes \ket{0} = \begin{pmatrix} 1 \\ 0 \end{pmatrix} \otimes \ket{0} = \begin{pmatrix} 1 \\ 0 \\ 0 \\ 0 \end{pmatrix}.
\end{equation}
The circuit can then be applied to the input vector:
\begin{equation}
    \frac{1}{\sqrt{2}}\begin{pmatrix} 1 & 0 & 1 & 0 \\ 0 & 1 & 0 & 1 \\ 0 & 1 & 0 & -1 \\ 1 & 0 & -1 & 0 \end{pmatrix} \begin{pmatrix} 1 \\ 0 \\ 0 \\ 0 \end{pmatrix}
    = \frac{1}{\sqrt{2}} \begin{pmatrix} 1 \\ 0 \\ 0 \\ 0 \end{pmatrix}
    = \frac{1}{\sqrt{2}} \begin{pmatrix} 1 \\ 0 \\ 0 \\ 1 \end{pmatrix}
    = \ket{\Phi^{+}}.
\end{equation}
Thus, from an initial input of $\ket{00}$ we have been able to prepare a Bell state using two simple quantum gates. The same procedure could be followed for all other possible input combinations of computational basis states to prepare the other Bell states.

The circuit above is one of the prototypical examples of quantum circuits and their usefulness. Later on, we shall utilise the tools learnt here to construct more complex circuits; these, in turn, will be used in our simulation.

\section{Noise}\label{sec:QC_noise}
No discussion of quantum computing would be complete without mention of the biggest challenge facing the field, noise. So far, we have dealt only with closed quantum systems, that is, systems which are completely isolated. Perfectly closed systems do not exist in practice; there will always be some level of interaction between a quantum system and its surroundings. Such systems are referred to as open quantum systems. Whenever a quantum system interacts with its surroundings, these interactions show up as noise in quantum information processing systems~\cite[Chapter~8]{Nielsen_Chuang_2010}. Random, uncontrollable physical processes in the qubit
control and measurement equipment, or in the local environment surrounding the quantum processor, are sources of noise that lead to decoherence~\cite{Krantz_2019}.

Decoherence is the term used to describe when a quantum system interacts with its environment, a quantum system will in general become rapidly and strongly entangled with a large number of environmental degrees of freedom~\cite{Schlosshauer_2019}. This entanglement strongly influences what can be observed when the quantum system is measured. In particular, quantum interference effects, with respect to certain physical quantities, become effectively suppressed, making them prohibitively difficult to observe in most cases of practical interest~\cite{Schlosshauer_2019}. As quantum computers require precise control of quantum principles such as superposition and entanglement, it is essential to minimise decoherence. Due to the extremely sensitive nature of controlling quantum systems, great measures are often taken to prevent decoherence as much as possible. A practical example of mitigating decoherence is found in superconducting qubit platforms, where the system is cooled to millikelvin temperatures using dilution refrigerators. For example, IBM's Goldeneye system was cooled to $\sim 25\,\mathrm{mK}$
in 2022~\cite{IBM_dilution}. This reduces thermal excitations due to the surrounding environment.

\begin{figure}[h]
\centering

\begin{subfigure}{0.3\textwidth}
    \centering
    \includegraphics[width=\linewidth]{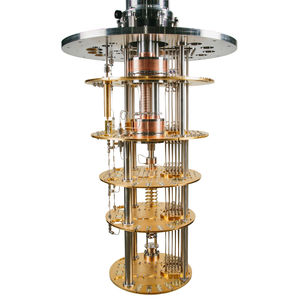}
    \caption{Dilution refrigerator.}
    \label{fig:dilution_refrigerator}
\end{subfigure}
\hspace{0.05\textwidth}
\begin{subfigure}{0.3\textwidth}
    \centering
    \includegraphics[width=\linewidth]{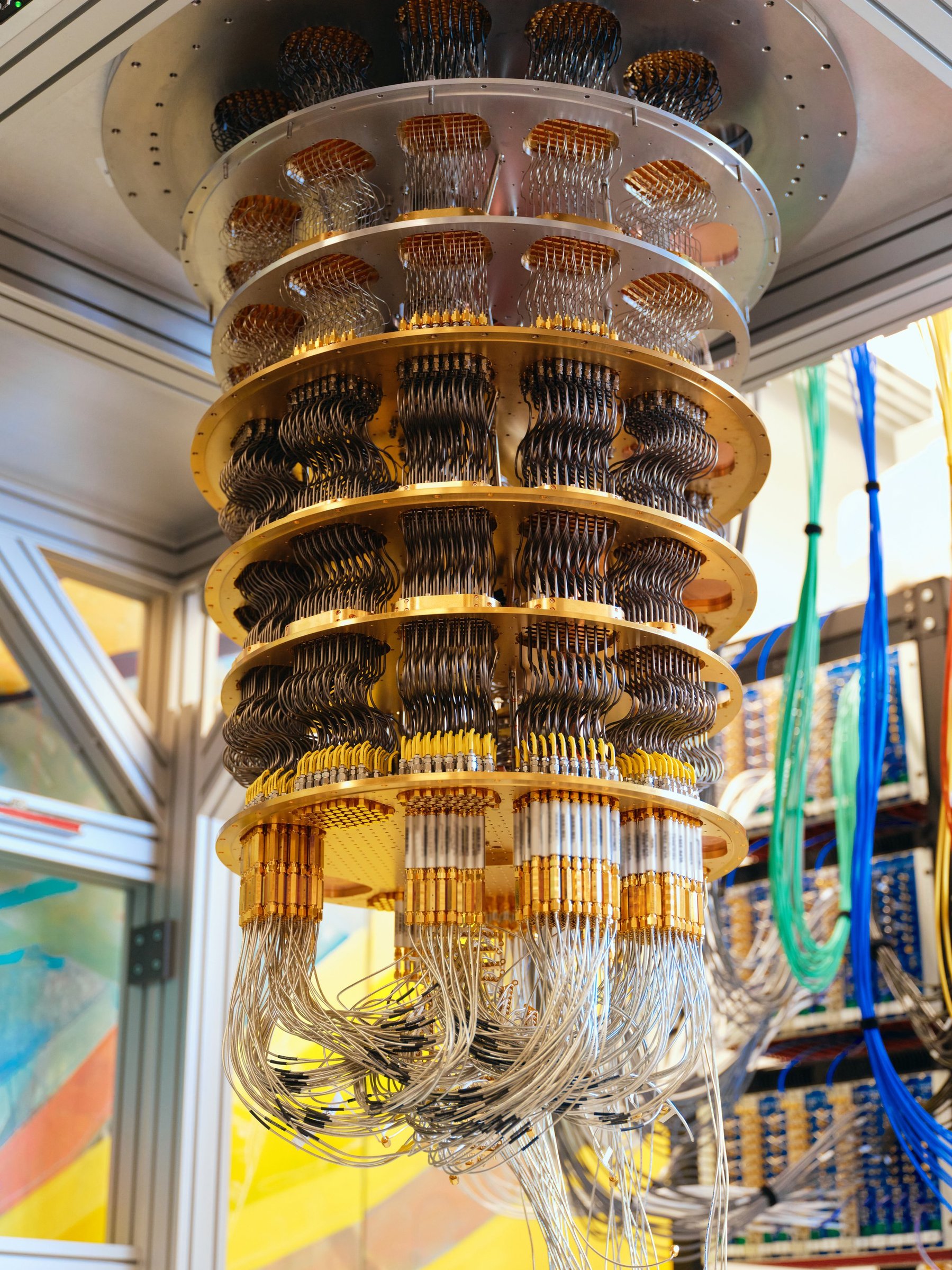}
    \caption{Dilution refrigerator with a wired quantum processor.}
    \label{fig:dilution_refrigerator_processor}
\end{subfigure}

\caption{Examples of dilution refrigerator systems used in superconducting quantum computing platforms. Images reproduced from Refs.~\cite{Quantum_design_dilution, Google_dilution}.}
\label{fig:dilution_refrigerators}
\end{figure}

We can categorise the various types of physical noise as follow~\cite{resch2021benchmarkingquantumcomputersimpact}:
\begin{itemize}
    \item \textbf{Interaction with Environment}. 
    As already mentioned, qubits need to be isolated from their environment in order to avoid any form of interaction, examples of possible sources of noise in this case are temperature, vibrations or magnetic fields.
    \item \textbf{Interaction with other qubit}. 
    Qubits can easily become entangled with each other, although we often want entanglement; the effect can still easily occur when it is not the desired outcome. The errors induced as a result of unintended qubit coupling are referred to as cross talk.
    \item \textbf{ Imperfect operation}. Imperfect application of quantum gates can generate incorrect quantum states~\cite{resch2021benchmarkingquantumcomputersimpact}. It is impossible for the fidelity (accuracy) of a quantum gate to reach $100\%$~\cite{10123493}.
    \item \textbf{Leakage}. Many quantum systems that are used as qubits actually have more than two possible states; thus, it is possible for a qubit to unintentionally enter one of these states, breaking the notion of a computational basis.
\end{itemize}

The errors induced by these different noise sources are by far the most problematic issue in quantum computing. Different processes have been developed, either practically or in theory, to deal with such errors. One of the most common techniques used in modern quantum computation is quantum error mitigation (QEM), which can be defined as the ability to minimise the noise-induced bias in expectation values on noisy hardware \cite{Cai_2023}. In practice, this means applying particular protocols to a noisy measurement result to attempt to reconstruct what the desired outcome should be, had there been no noise. Later on, we shall implement partial QEM to mitigate some sources of error.

Whilst not the main focus of this work, no discussion of quantum computing would be complete without the mention quantum error correction (QEC). QEC, unlike QEM, operates directly on quantum states, rather than the expectation values that QEM targets through post processing. The aim of QEC is to allow for quantum information to be redundantly encoded by entangling it across an expanded system of qubits \cite{Roffe_2019}, this entangled system is called a logical qubit. The hope of QEC is that errors within the logical qubit can be detected in real time and corrected. Whilst a very promising and elegant idea, current devices lack the scale and fidelity required for fully fault-tolerant quantum computation. For this reason we shall be focusing on QEM for the rest of this work.

\chapter{The Variational Quantum Eigensolver}

\label{sec:VQE}
\section{Introduction}\label{sec:VQE_intro}

The variational quantum eigensolver (VQE) is a hybrid quantum-classical algorithm that uses the variational principle of quantum mechanics to estimate the energies of a quantum system. The hybrid nature of the VQE makes it well suited to NISQ devices. NISQ typically refers to quantum computers containing around $50-500$ physical qubits (Google's Willow chip: $105$ qubits, IBM Nighthawk: $120$~\cite{{IBM_qubit_count}, {Google_qubit_count}}). The VQE is especially useful in solving well known physics problems, such as the TFIM, where we can leverage our physical knowledge of the problem to enhance the design and implementation of a given algorithm.

The VQE can be split into several steps, which proceed as follows~\cite{Tilly_2022}:
\begin{enumerate}
    \item Prepare an initial parametrised circuit $\ket{\psi(\boldsymbol{\theta})}$ which is referred to as an ansatz.
    \item Calculate the energy expectation value
    $E(\boldsymbol{\theta}) = \expval{\hat{H}}{\psi(\boldsymbol{\theta})}$.
    \item Using a classical optimiser vary the parameters $\boldsymbol{\theta}$ in order to minimise $E(\boldsymbol{\theta})$.
    \item Repeat steps $2$ and $3$ until the energy converges either to a local or global minimum.
\end{enumerate}

In preparing the ansatz we are restricting the search space to a parametrised subset of the Hilbert space. The optimiser then must search through this parameter space using the energy expectation values as a starting point. Finally, after a sufficient number of iterations, a global minimum should have been reached, providing an accurate and efficient estimate of the ground-state energy. This is the essence of the VQE process and we shall use the remainder of this chapter to explain this in further detail.
\begin{figure}[H]
    \centering
    \includegraphics[width=0.7\linewidth]{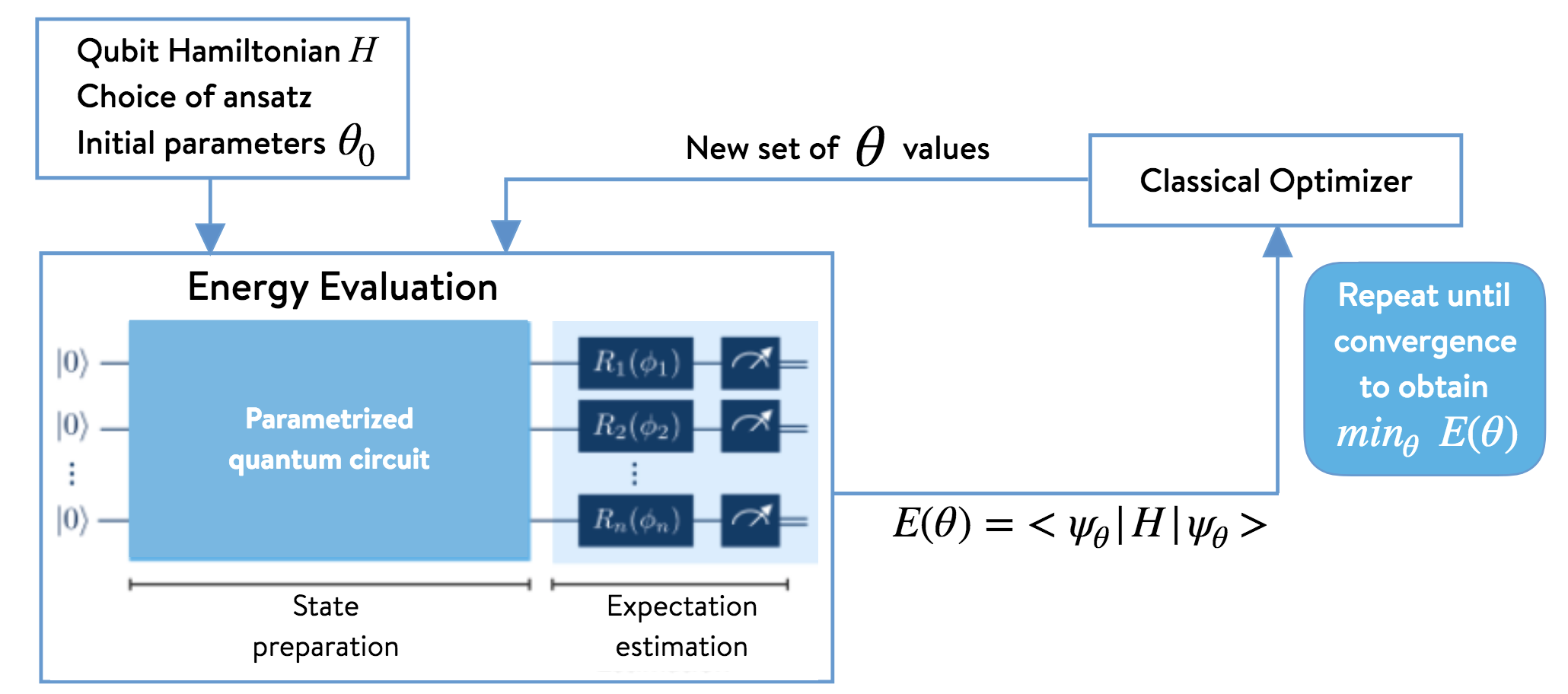}
    \caption{VQE process (taken from \href{http://openqemist.1qbit.com/docs/vqe_microsoft_qsharp.html}{OpenQEMIST documentation}).}
    \label{fig:vqe_process}
\end{figure}

\section{Hamiltonian representation}\label{sec:ham_rep}
Before we can begin explaining each of the key VQE steps in detail, we should mention how quantum computers understand Hamiltonians. While classical computers require numerical data to be encoded in binary, quantum computers require observables to be decomposed into operators that can be measured on qubits. Qubit registers on quantum computers can only measure observables expressed in a
Pauli basis~\cite{Tilly_2022}, specifically the $z$-basis; thankfully, the TFIM Hamiltonian~\eqref{eq:hamiltonian_TFIM} is already expressed in a Pauli basis. In order to calculate the expectation values required for the VQE, we must decompose the Hamiltonian into a sum of Pauli strings consisting only of possible values $\{I,X,Y,Z\}$, where each letter represents one of the Pauli operators. Once expressed in this form, a quantum computer can evaluate expectation values using the parametrised state $\ket{\psi(\boldsymbol{\theta})}$.

We will demonstrate how this works by considering a two-qubit spin chain. The summations within~\eqref{eq:hamiltonian_TFIM} take the form:
\begin{equation}
  \begin{aligned}
    \sum_{i=1}^{2} \sigma_{i}^z \sigma_{i+1}^z 
    &= \sigma_{1}^z \sigma_{2}^z + \sigma_{2}^z \sigma_{1}^z \\
    \sum_{i=1}^{2} \sigma_{i}^x &= \sigma_{1}^x + \sigma_{2}^x.
\end{aligned}  
\end{equation}
Now, recalling how we represent Pauli operators within a one-dimensional lattice, we have $\sigma_{1}^z = \sigma^z \otimes I$. The operator notation is then replaced with the Pauli string notation, and we are left with the following Hamiltonian representation:
\begin{equation}
    \hat{H} = -J(ZZ+ZZ) - h(XI + IX).
\end{equation}
Each of these terms can be measured individually on a quantum computer, with appropriate basis rotations applied when necessary. The final expectation value is then obtained by combining the measured results according to their corresponding coefficients. Note that these strings may often be reversed depending on the specific software used; for example, Qiskit orders Pauli strings from right to left.

\section{Ansatz selection and preparation}\label{sec:ansatz_prep}
The goal of the VQE is the minimisation of some cost function. To adapt this minimisation task to a quantum computer, one must start by defining a so-called ansatz wave function that can be implemented on a quantum device as a series of quantum
gates~\cite{Tilly_2022}. These quantum gates are given an initial randomised set of parameter values $\boldsymbol{\theta}$. The ansatz is used to produce a trial state, from which an energy expectation value is calculated before the parameters are optimised for the next iteration. 

Therefore, it is crucial we select an appropriate ansatz. If the initial ansatz is insufficient, the state it first generates will most likely be too far away from a desired result. This, in turn, can cause the optimiser to get stuck in a local minimum and not reach an appropriate estimate of the ground state energy. To successfully find an optimal solution, the ansatz must be both expressive and trainable~\cite{Holmes_2022}.

The expressibility defines the ability of the ansatz to span a large class of states in the Hilbert space~\cite{Tilly_2022}, in other words, our ansatz must be sufficiently large that a minimum can be successfully reached. Whilst ansatz expressibility is key to finding a minimum, it should be noted that as circuit depth increases we will observe diminishing returns. This is because both the cost and error will grow as circuits increase, eventually this will outweigh the benefits of increased expressivity. The act of balancing expressivity versus cost is a crucial step when selecting a model's ansatz. A good ansatz must be sufficiently expressive to guarantee that it can appropriately approximate the ground state wave function, however, it must not be so expressive that it renders the search for the target state intractable~\cite{Holmes_2022}. These considerations become particularly  meaningful as the number of qubits grows; in this case, one must take careful consideration of ansatz depth, as this largely determines the noise profile of the circuit.

Ansatz trainability describes the practical ability of the ansatz to be optimised using techniques tractable on quantum devices~\cite{Tilly_2022}. Specifically, the trainability is the process of being able to find the best set of parameters for the ansatz. If our ansatz is untrainable, we will most likely encounter the infamous barren plateau problem. The barren plateau problem occurs when an optimiser is searching an essentially flat subspace of the Hilbert space; as a result, no minimum can be found. McClean et al.~\cite{McClean2018} showed that randomly initialised parameters, indicative of variational quantum algorithms, can lead to barren plateaus in the Hilbert subspace. As a consequence, it is essential that the ansatz be sufficiently trainable in order to optimise these randomly initialised parameters.

Having established the key properties required of a suitable ansatz, we now turn to how such ans\"atze are constructed in practice. Generally, the structure of the ansatz can be split into two types: fixed or adaptive. Fixed structure ans\"atze are initialised at the beginning of the VQE process and remain unchanged (aside from the value of their parameters) throughout the optimization. In contrast to this, adaptive ansatz grow iteratively throughout the optimisation process by adding new operators~\cite{Tilly_2022}. For this work, we will be focusing on a specific fixed-structure ansatz known as the hardware efficient ansatz (HEA).

The HEA was originally introduced as an ansatz tailored to the specific basis gates of a quantum device. The main objective of HEA is to mitigate the effect of hardware noise by using native gates~\cite{Leone2024practicalusefulness}. Whilst several variations of the HEA exist, they all follow a similar structure: the ansatz is constructed by repeating blocks of single-qubit parametrised rotation gates and ladders of entangling gates~\cite{Tilly_2022}.

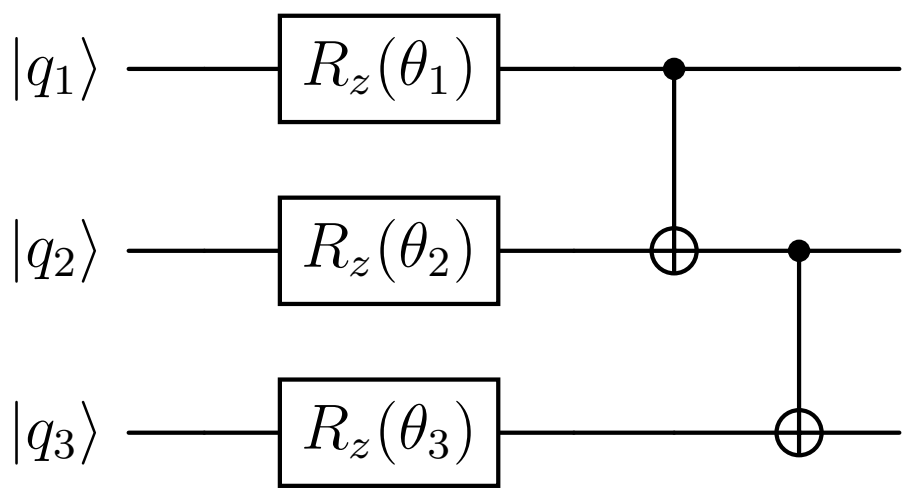
\begin{figure}[H]
    \centering
    \begin{quantikz}
        \lstick{$\ket{q_1}$}&&\gate{R_z(\theta_1)}&&\ctrl{1}&& \\
        \lstick{$\ket{q_2}$}&&\gate{R_z(\theta_2)}&&\targ{1}&\ctrl{1} & \\
        \lstick{$\ket{q_3}$}&&\gate{R_z(\theta_3)}&&&\targ{1}&
    \end{quantikz}
    \caption{Example of a HEA containing one set of rotation and entanglement gates with 3 randomised parameters.}
    \label{fig:HEA_example}
\end{figure}

A major advantage of the HEA is its versatility, this allows the ansatz to be both easy to construct as well as expressive. The HEA is particularity useful for small-scale quantum simulations where the qubit count is kept relatively low. This choice is motivated by the work of Ref.~\cite{Cerezo_2021}, where it was shown that shallow HEAs can be resistant to barren plateaus, helping to preserve trainability. It is worth noting, the HEA often struggles, particularly for larger systems, due to the fact that it must span a very large
portion of the Hilbert space to guarantee that an accurate enough representation of the ground state wave function can be produced~\cite{Tilly_2022}. Although this is a drawback, the simplicity of the HEA makes it a suitable candidate for the simulations carried out in this work.

\section{Calculating expectation values}\label{sec:calc_expec_vals}

Having initialised a parametrised ansatz, we now move to the next stage in the VQE pipeline, namely the calculation of energy expectation values. In quantum mechanics, the energy expectation value is given by
\begin{equation}
    \label{eq:gen_expec_val}
    E(\boldsymbol{\theta}) = \expval{\hat{H}}{\psi(\boldsymbol{\theta})}.
\end{equation}
Now, we recognise $\psi(\boldsymbol{\theta})$ as our initial parametrised circuit. In order to calculate the expectation value~\eqref{eq:gen_expec_val} on a quantum computer, it will be essential to use the Hamiltonian representation introduced in~Sec.~\ref{sec:ham_rep}. As previously mentioned, quantum computers can only understand Hamiltonians expressed as sums of Pauli strings. If $P_a$ is some general Pauli string, then the Hamiltonian can be expressed as
\begin{equation}
    \hat{H} = \sum_{a=1}^{P}\omega_{a}P_a,
\end{equation}
where $\omega_a$ is some weighting coefficient and $P$ is the total number of Pauli strings.

Using this representation, we can re-express~\eqref{eq:gen_expec_val} as
\begin{equation}
    E(\boldsymbol{\theta}) = \sum_{a=1}^{P}\omega_{a}\expval{P_a}{\psi(\boldsymbol{\theta})}.
\end{equation}
Whilst this expression is sufficient for a quantum computer, in order for us to truly appreciate what is happening it is worth re-formulating the ansatz. Recalling~Sec.~\ref{subsec:circuits}, quantum circuits consist of gates represented by unitary operators that act on qubits. Generally, the qubits at the beginning of a circuit are initialised to $\ket{0}$. For an $N$-qubit system, this  initialisation can be represented as $\ket{0}^{\otimes N} \equiv \ket{\boldsymbol{0}}$~\cite{Tilly_2022}. Thus, denoting the general unitary operator of a circuit as $U(\boldsymbol{\theta})$, the ansatz can be expressed as
\begin{equation}
    \ket{\psi(\boldsymbol{\theta})} = U(\boldsymbol{\theta})\ket{\boldsymbol{0}}.
\end{equation}

Bringing everything together, we are led to the following expression for the energy expectation value:
\begin{equation}
    E(\boldsymbol{\theta}) = \sum_{a}^{P}\omega_{a}\expval{U(\boldsymbol{\theta})^{\dagger}P_{a}U(\boldsymbol{\theta})}
    {\boldsymbol{0}}.
\end{equation}
The summands correspond to the expectation value of some Pauli string  $P_{a}$, which can be computed on a quantum device, whilst the summation and subsequent minimisation are computed on a classical computer~\cite{Tilly_2022}. In practice, the expression above is evaluated term by term on the quantum device. The parametrised circuit $U(\boldsymbol{\theta})$  is executed to prepare $\ket{\psi(\boldsymbol{\theta})}$, and any non-$Z$ factors in $P_a$ are first rotated into the computational basis before measurement. By repeating this process over many circuit executions (shots), we are able to estimate the expectation value from the resultant measurement distribution. This process of performing a finite number of shots introduces an additional source of noise to the VQE, affecting the overall noise landscape.

\section{Choosing an optimiser}\label{sec:optimiser}

The final stage in the VQE process, before iterating, is to use a classical optimiser to minimise the energy expectation value by varying the parameters until convergence. In general, this requires sampling the expectation value of the Hamiltonian several times for a given parameter set in the ansatz~\cite{Tilly_2022}. Although this is a classical process, the specific optimisation required for the VQE faces additional challenges; the noise profile of NISQ devices disturbs the landscape of the objective function (the energy) and can be detrimental to the convergence of optimization~\cite{Tilly_2022}.

Due to the noisy nature of quantum computing, the objective function landscapes will often reflect such noise, leading to gradients that cannot be computed analytically. This is a major drawback of local, gradient-based optimisers such as BFGS~\cite{doi:10.1137/0916069} or L-BFGS-B~\cite{10.1145/279232.279236}. In order for this class of optimisers to compute gradients on noisy devices, they must employ compatible
gradient evaluation methods, which in turn creates noisy gradient evaluations~\cite{kerfoot2025quantumvariationalmethodssupersymmetric}.

Naturally, the next type of optimisers to consider would be local, gradient-free optimisers such as COBYLA~\cite{Powell1994} or COBYQA~\cite{rago_thesis}. These methods circumvent the need to calculate gradients by constructing local approximations
of the objective function within a trust region~\cite{kerfoot2025quantumvariationalmethodssupersymmetric}, COBYLA uses linear approximations, whilst COBYQA uses quadratic. Their relative simplicity and ability to handle noisy function evaluations make such
optimisers particularly well suited to VQE applications~\cite{carreras2026limitations}. Despite this, gradient free optimisers still encounter problems in the VQE; for example, it was shown in~Ref.~\cite{ortiz2021entanglement} that gradient free methods are still susceptible to the barren plateau problem introduced previously. Additionally,  the local nature of these methods can become an issue when the energy landscape is noisy, with it being common for these methods to get stuck in local minima~\cite{kerfoot2025quantumvariationalmethodssupersymmetric}. These considerations reinforce the need to construct an efficient ansatz.

In addition to local optimisers, several other classes of optimiser may be used. These include stochastic methods, which estimate properties of an expectation value by sampling functions that depend on random variables~\cite{Tilly_2022}. Whilst such methods seem promising, stochastic differential evolution optimisers have, for example, been shown to be effective in avoiding local minima in variational quantum algorithms~\cite{kerfoot2025quantumvariationalmethodssupersymmetric, Fa_lde_2023}. However, the number of circuit evaluations required for such methods can become large. For small-scale simulations, such as those conducted in this work, this scaling of cost is inefficient and unnecessary. As a result, we shall be considering only local optimisers that are either gradient-free or gradient-based.

\chapter{Results}
\label{sec:results}
\section{Initialisation}\label{sec:resukts_init}

We begin the VQE simulation of the TFIM by initialising the algorithmic setup. All computational calculations will be carried out using the IBM Quantum Qiskit~\cite{qiskit2024} framework in Python. To initially model noise we use an IBM Quantum fake backend, these are previously operational quantum computers whose noise properties we use to simulate real hardware. The specific backend we use is \texttt{FakeManilaV2}~\cite{IBM_retired_QPUs}, which was a $5$-qubit system that was retired in $2023$. Later, when simulating using real hardware we use \texttt{ibmq\_fez} (version $1.3.37$)~\cite{IBM_fez}, which is a $156$ qubit device operating on a Heron r$2$ quantum processing unit (QPU). 

We perform all calculations using $10$ independent VQE runs with $1000$ shots per iteration. For expectation value calculations, we employ the Qiskit \texttt{EstimatorV2} primitive~\cite{IBM_estimator}, which enables efficient evaluation of the Hamiltonian expectation values required for the VQE cost function. In a noisy regime, the estimator is used to estimate the expectation value using a finite number of shots. For measurements involving observables such as magnetisation, we use the \texttt{SamplerV2} primitive~\cite{IBM_sampler}, which returns measurement outcome distributions from which expectation values can be reconstructed.

\begin{figure}
    \centering
    \begin{subfigure}{0.3\textwidth}
        \centering
        \includegraphics[width=\linewidth]{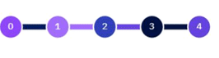}
        \label{fig:left}
    \end{subfigure}
    \hspace{0.05\textwidth}
    \begin{subfigure}{0.3\textwidth}
        \centering
        \includegraphics[width=\linewidth]{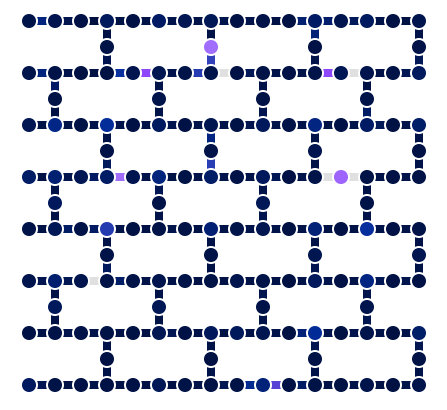}
        \label{fig:right}
    \end{subfigure}

    \caption{Left: IBM Quantum Manila qubit layout (taken from Zhang et al.~\cite{zhang2021implementation}). Right: IBM Quantum Fez qubit layout (taken from~\cite{IBM_fez}).}
    \label{fig:fez_manila_qubit_layout}
\end{figure}

We shall be considering a four-qubit system, inducing the space $\mathcal{H}_{2}^{\otimes4}$ with dimension 16. We choose four qubits as they represent a interesting toy model, which can be easily compared analytically; a future direction would be to test the optimised VQE configuration using a larger number of qubits and determine whether similar levels of accuracy are maintained. The three VQE components requiring initialisation are: ansatz, parameters and optimiser. We construct the HEA~\footnotemark ~using the \texttt{n\_local} class~\cite{IBM_n_local}, consisting of three rotation layers of parametrised $R_y$ rotation gates and two layers of CNOT entanglement gates. The entanglement strategy used is circular, which entangles adjacent qubits as well as the first and last qubit~\cite{IBM_n_local}, thereby reflecting the use of periodic boundary
conditions.  

\footnotetext{Note that whilst this ansatz follows the structure of a HEA, it can't truly be considered a HEA as the native gates for FakeManilaV2 contain $R_z$ gates instead of $R_y$.}

\begin{figure}[h]
    \centering
    \includegraphics[width=0.9\linewidth]{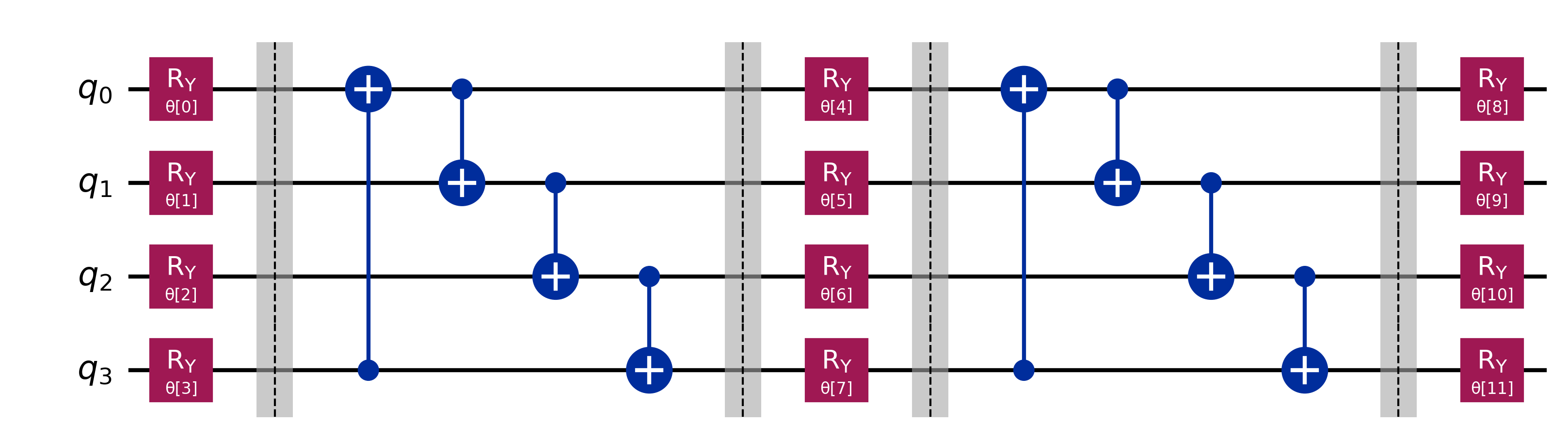}
    \caption[Initial ansatz.]{Initial ansatz.}
    \label{fig:init_ansatz}
\end{figure}

As can be seen from~{Fig.}~\ref{fig:init_ansatz}, we must initialise $12$ random parameters $\theta \in [0,2\pi]$, which we can think of as initial angles on the Bloch sphere (see~{Fig.}~\ref{fig:Bloch_sphere}). For the optimiser, we begin using the gradient free COBYLA with the maximum number of iterations per optimiser call set to $750$. We summarise the full initialisation in~{table}~\ref{tab:vqe_init}.

\begin{table}[H]
    \centering
    \begin{tabular}{ll}
    \toprule
    \textbf{Component} & \textbf{Specification} \\
    \midrule
    System size & 4 qubits  \\
    Ansatz type & Hardware-efficient ansatz (HEA) \\
    Rotation gates & $R_y$ \\
    Entanglement gates & CNOT ($CX$) \\
    Entanglement strategy & Circular \\
    Number of layers & 3 rotation layers, 2 entangling layers \\
    Parameters & 12 parameters, $\theta \in [0, 2\pi]$ \\
    Optimiser & COBYLA \\
    Maximum optimiser iterations & $750$ \\
    Backend (noise model) & \texttt{FakeManilaV2} \\
    Backend (hardware) & \texttt{ibmq\_fez} (version 1.3.37) \\
    \bottomrule
    \end{tabular}
    \caption{Summary of the VQE initialisation and ansatz configuration.}
    \label{tab:vqe_init}
\end{table}

\begin{figure}[h]
    \centering
    \includegraphics[width=1.0\linewidth]{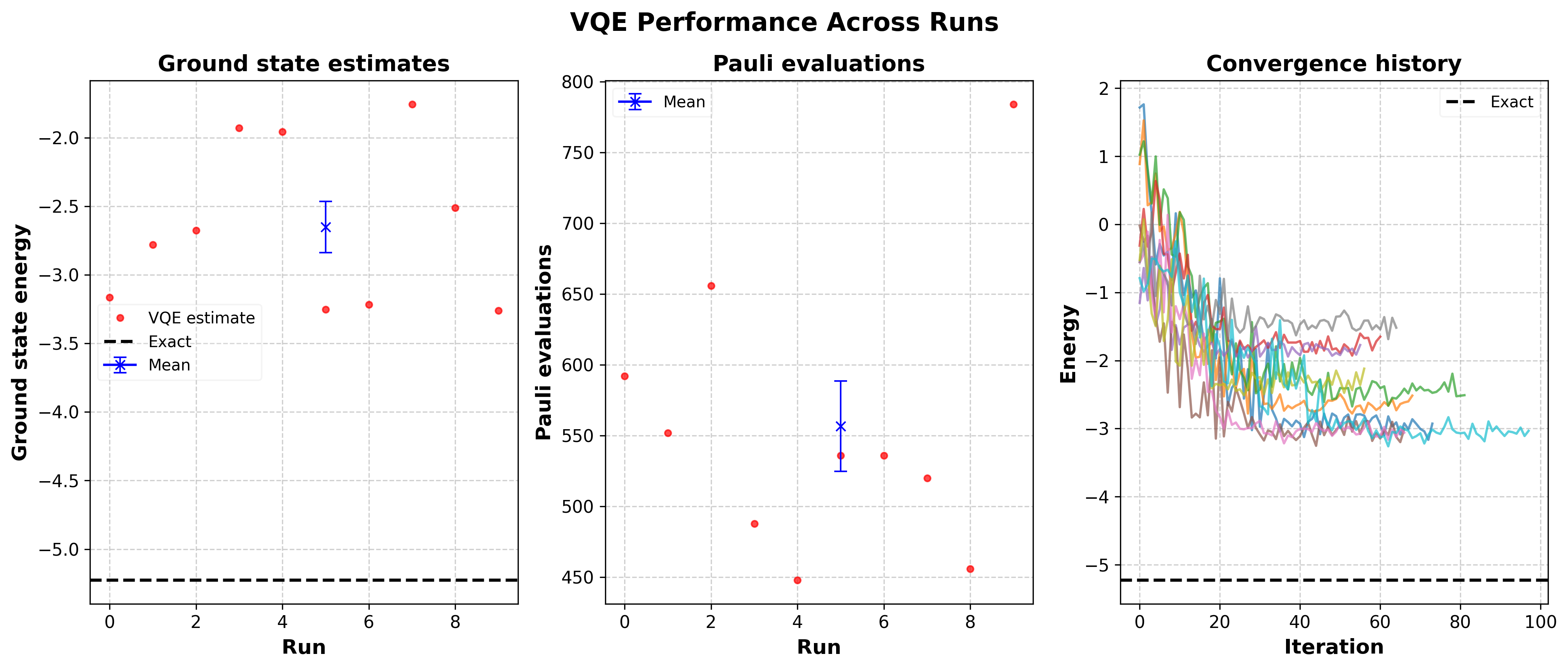}
    \caption{Results for the initialised VQE setup using 10 runs with 1000 shots}
    \label{fig:init_VQE_results}
\end{figure}

Using this initial setup, with $h=1$ and $J=1$, we obtain our first estimate of the ground state energy using the \texttt{FakeManilaV2} backend. We then compare this initial estimate with the exact value, $E_{0}^{\text{exact}} = -5.23$~\footnote{Recall from~{Sec.}~\ref{sec:ED} this was $E_0/J$ to ensure dimensionless units. As $J=1$ we omit the denominator.}. The three graphs we will use for analysis are:
\begin{enumerate}
    \item Energies for direct comparison.
    \item Number of Pauli string evaluations to gauge cost. 
    \item Energy histories to understand convergence.
\end{enumerate}
We present our results in~{Fig.}~\ref{fig:init_VQE_results} and~{table}~\ref{tab:init_vqe_results}.

\begin{table}[h]
    \centering
    \begin{tabular}{ccc c ccc}
        \toprule
        Run & $E_{0}^{\mathrm{VQE}}$ & $N_{\mathrm{pauli}}$
        & \hspace{0.4cm} &
        Run & $E_{0}^{\mathrm{VQE}}$ & $N_{\mathrm{pauli}}$ \\
        \midrule
        1 & $-3.17$ & $592$ & & 6  & $-3.25$ & $536$ \\
        2 & $-2.78$ & $552$ & & 7  & $-3.22$ & $536$ \\
        3 & $-2.68$ & $656$ & & 8  & $-1.76$ & $520$ \\
        4 & $-1.93$ & $488$ & & 9  & $-2.51$ & $456$ \\
        5 & $-1.96$ & $448$ & & 10 & $-3.26$ & $784$ \\
        \bottomrule
    \end{tabular}
    \caption{Initial VQE energy estimates and corresponding number of Pauli string evaluations over ten runs.}
    \label{tab:init_vqe_results}
\end{table}

Whilst this initialisation does always converge to a value, the mean ground state energy across the ten runs is 
\begin{equation}
    \label{eq:init_VQE_E0}
    \overline{E_{0}^{VQE}} = -2.65 \pm 0.19,
\end{equation} 
where the error has been calculated as the standard error on the mean. Using~\eqref{eq:init_VQE_E0}, we can estimate the relative error:
\begin{equation}
    \epsilon_{\mathrm{rel}} = \frac{|\overline{E_{0}^{\mathrm{VQE}}} - E_{0}^{\mathrm{exact}}|}{|E_{0}^{\mathrm{exact}}|} \times 100 =
    \frac{|-2.65 - (-5.23)|}{|-5.23|} \times 100
    \approx 49\%.
\end{equation}
The corresponding mean number of Pauli string evaluations for this initialisation is
\begin{equation}
    \overline{N}_{\mathrm{Pauli}} = 557 \pm 32. 
\end{equation}

We would like to have a more accurate energy estimate. To achieve this, we will perform different experiments on the VQE setup, with the aim of improving the resulting ground-state estimate.

\section{Optimising the VQE}\label{sec:optimise_VQE}
During this work, we performed numerous experiments of VQE setups in order to try and find the most optimal configuration. In this chapter, we present some of the more interesting results of these experiments; the remaining results can be found in Appendix~\ref{Apdx:extra_results}. For each experiment, we observed the effects of changing a given VQE component whilst keeping all other components the same. In each case, we performed ten independent runs per configuration and analysed the mean energies, mean Pauli evaluations, and the convergence history of the run with the lowest number of iterations. This choice is appropriate here because the aim of these preliminary experiments was to identify a well-performing ansatz and optimisation setup for use in later simulations, rather than to characterise the worst-case behaviour of the algorithm.

\subsection{Classical optimisers}
We first decided to test the classical component of our VQE setup, the optimiser. For this, we tested two gradient-based optimisers, BFGS and L-BFGS-B, in addition to two gradient-free optimisers, COBYLA and COBYQA.

\begin{table}[h]
    \centering
    \label{tab:noisy_optimizer_results}
    \begin{tabular}{ccccc}
        \toprule
        Optimiser & $\overline{E_{0}^{VQE}}$ & $\Delta\overline{E_{0}^{VQE}}$ & $\overline{N_{pauli}}$ & $\Delta \overline{N_{pauli}}$ \\
        \midrule
        COBYLA   & $-2.00$ & $0.20$ & $506$  & $22$ \\
        COBYQA   & $-3.36$ & $0.070$ & $1000$  & $27$ \\
        BFGS     & $-0.154$ & $0.30$ & $2590$ & $150$ \\
        L-BFGS-B & $-0.000$  & $0.31$ & $2560$ & $150$ \\
        \bottomrule
    \end{tabular}
    \caption{Summary of noisy VQE results for different classical optimisers. Energy and Pauli evaluation uncertainties are reported as standard errors of the mean.}
\end{table}

\begin{figure}[h]
    \centering
    \includegraphics[width=1.0\linewidth]{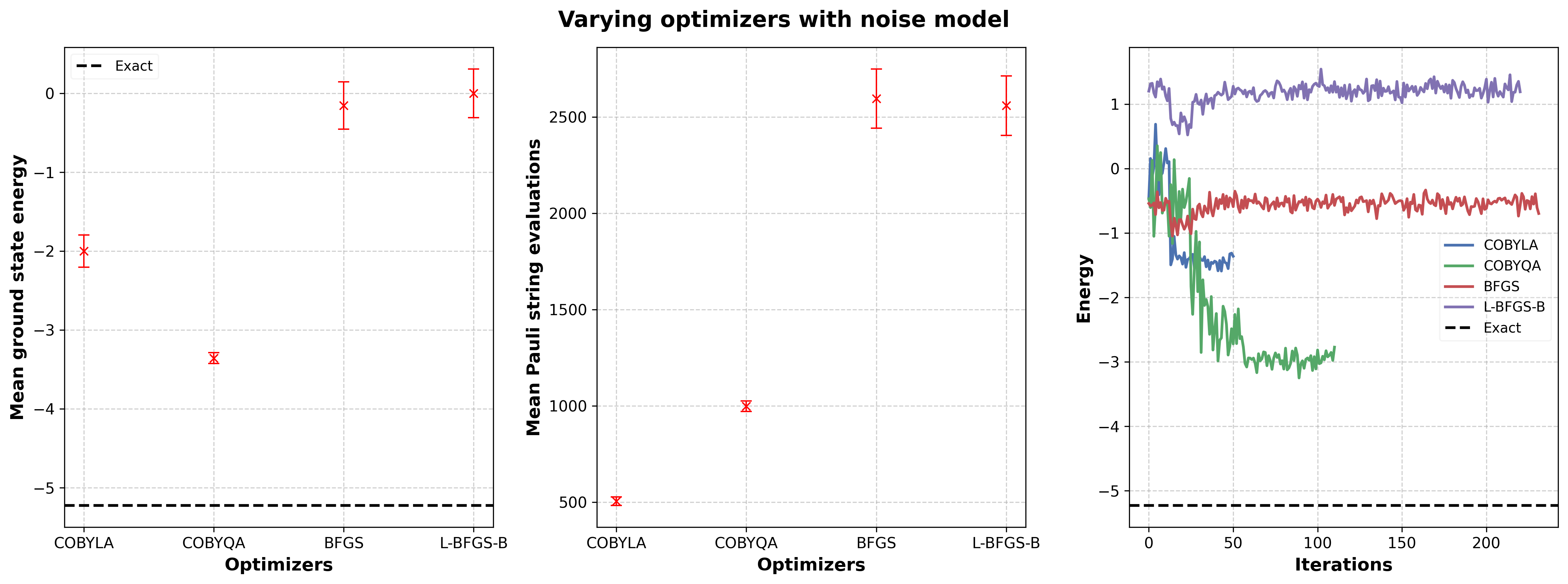}
    \caption{Results for varying the optimiser while keeping all other components fixed}
    \label{fig:vary_opt}
\end{figure}

Firstly, we observe both gradient-based optimisers performing poorly in terms of accuracy and cost. This is most likely a consequence of the noisy parameter space, which as discussed in~{Ref.}~\ref{sec:optimiser}, leads to non-analytic gradient computations. Both gradient-free optimisers perform better, with COBYQA achieving the best accuracy; we believe this could be due to COBYQA using quadratic approximations instead of the linear approximations used by COBYLA. Although COBYQA does have a slightly higher cost, we believe the accuracy gained through COBYQA justifies this gain in cost.

\subsection{Entanglements}
Moving to the quantum component of the VQE, we first investigate the effect different entanglement types on the ground-state energy. We tested each of the built in entanglement types for the \texttt{n\_local} class; see~{Ref.}~\cite{IBM_n_local} for specific details of each.

\begin{table}[h]
    \centering
    \label{tab:noisy_entanglement_results}
    \begin{tabular}{ccccc}
        \toprule
        Entanglement & $\overline{E_{0}^{VQE}}$ & $\Delta\overline{E_{0}^{VQE}}$ & $\overline{N_{pauli}}$ & $\Delta \overline{N_{pauli}}$ \\
        \midrule
        Circular        & $-2.57$ & $0.23$ & $546$ & $23$ \\
        Linear          & $-3.62$ & $0.18$ & $522$ & $21$ \\
        Reverse linear  & $-3.51$ & $0.20$ & $540$ & $29$ \\
        Pairwise        & $-3.67$ & $0.14$ & $536$ & $29$ \\
        SCA             & $-2.93$ & $0.27$ & $573$ & $30$ \\
        \bottomrule
    \end{tabular}
    \caption{Summary of noisy VQE results for different entanglement strategies. Energy and Pauli evaluation uncertainties are reported as standard errors of the mean.}
\end{table}

\begin{figure}[h]
    \centering
    \includegraphics[width=1.0\linewidth]{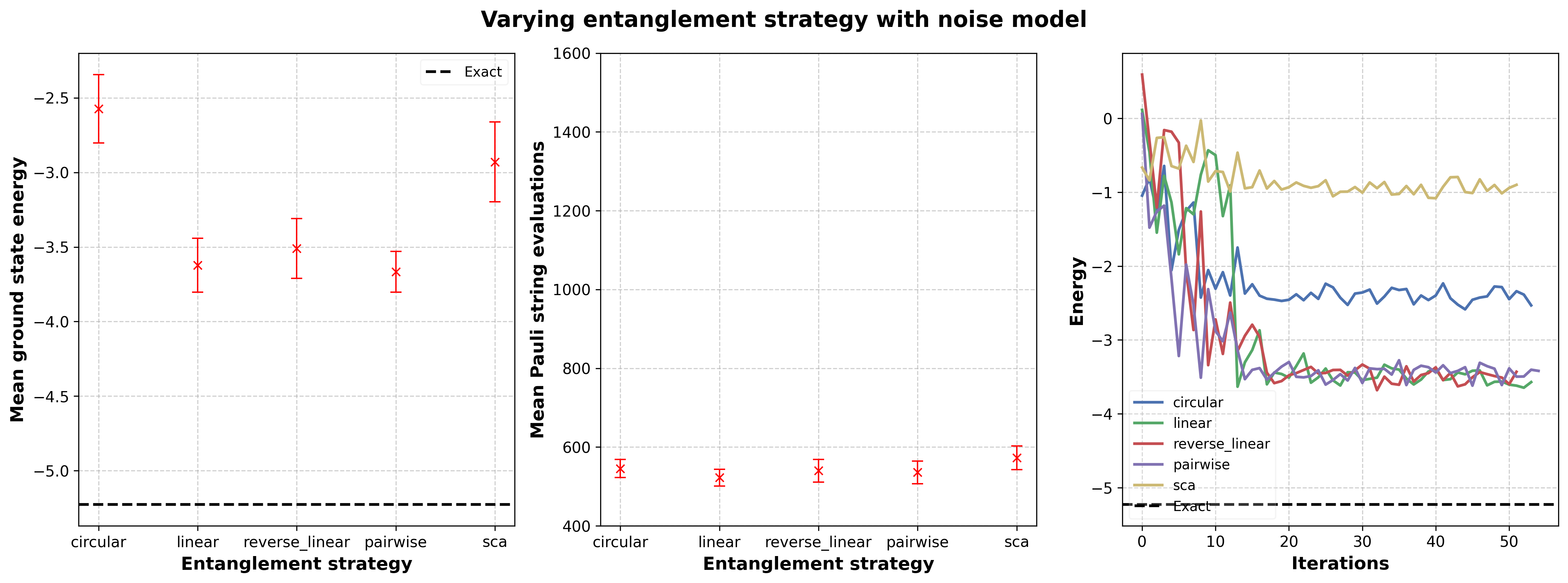}
    \caption{Results for varying the entanglement types whilst keeping all other components fixed}
    \label{fig:vary_ent_types}
\end{figure}

Interestingly, we observe that circular actually performs the worst, this was a surprise as we felt that this entanglement best matched the periodic boundary conditions of the Hamiltonian. This poor performance, we believe, is due to the additional work required for the `long' entanglement of the first and last qubit. In particular, this result demonstrates the common trade-off in quantum computation between physical accuracy and computational cost, with physically motivated ansatz not always leading to optimised results on NISQ devices.

When evaluating energy vs. cost, we concluded that linear entanglement was the most efficient. Unlike circular entanglement, linear entanglement only entangles adjacent qubits, which in our case removes the CNOT between qubit $0$ and $3$.

\subsection{Rotation gates}

Additionally, we wanted to understand the effects of increasing the ansatz expressivity. One way of increasing expressivity is by increasing the number of gates within the circuit. To maintain the structure of a HEA, we test only parametrised rotation gates.

\begin{table}[h]
    \centering
    \label{tab:noisy_rotation_results}
    \begin{tabular}{ccccc}
        \toprule
        Rotation block & $\overline{E_{0}^{VQE}}$ & $\Delta\overline{E_{0}^{VQE}}$ & $\overline{N_{pauli}}$ & $\Delta \overline{N_{pauli}}$ \\
        \midrule
        $R_y$              & $-2.31$ & $0.20$ & $540$  & $29$ \\
        $R_y, R_x$         & $-2.47$ & $0.13$ & $1100$ & $66$ \\
        $R_y, R_z$         & $-2.47$ & $0.20$ & $1150$ & $58$ \\
        $R_z, R_x$         & $-2.06$ & $0.10$ & $1090$ & $35$ \\
        $R_y, R_x, R_z$    & $-2.45$ & $0.17$ & $1570$ & $50$ \\
        \bottomrule
    \end{tabular}
    \caption{Summary of noisy VQE results for different ansatz rotation blocks. Energy and Pauli evaluation uncertainties are reported as standard errors of the mean.}
\end{table}

\begin{figure}[h]
    \centering
    \includegraphics[width=1.0\linewidth]{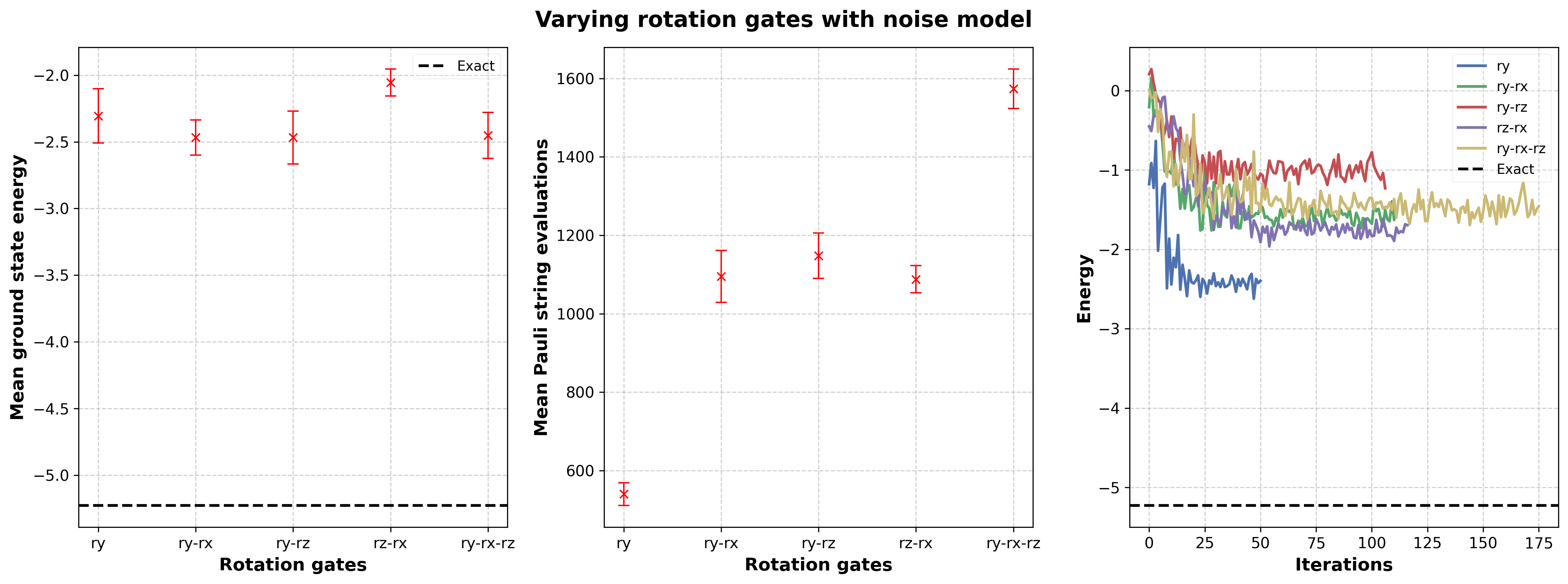}
    \caption{Results for varying the rotation gates whilst keeping all other components fixed}
    \label{fig:vary_rot}
\end{figure}

From~{Fig.}~\ref{fig:vary_rot}, we observe all gate variations give similar results for the energy estimates. Specifically, the mean energy results seem to suggest there is no dependence on the selection of rotation gates. Despite this, the cost plot does indicate a drawback in increasing expressivity, with the largest cost result coming from the most expressive circuit. This trade-off between accuracy and cost was discussed in Sec.~\ref{sec:ansatz_prep} with emphasis on systems with large qubit counts; however, even for our small four-qubit system, we observe the impact that expressivity can have on overall cost.

\subsection{Final optimised VQE configuration}\label{sec:best_VQE}
As a result of these experiments, and others, we can now introduce an optimised VQE setup. All the relevant changes are detailed in~{Table}~\ref{tab:vqe_changes}.

\begin{table}[h]
    \centering
    \begin{tabular}{lll}
    \toprule
    \textbf{Component} & \textbf{Initial specification} & \textbf{Current specification} \\
    \midrule
    Rotation gates & $R_y$ & $R_y$, $R_z$ \\
    Entanglement strategy & Circular & Linear \\
    Number of layers & 3 rotation, 2 entangling & 4 rotation, 3 entangling \\
    Parameters & 12 & 32 \\
    Optimiser & COBYLA & COBYQA \\
    \bottomrule
    \end{tabular}
    \caption{Summary of the changes made to the VQE initialisation.}
    \label{tab:vqe_changes}
\end{table}

\begin{figure}[H]
    \centering
    \includegraphics[width=\textwidth]{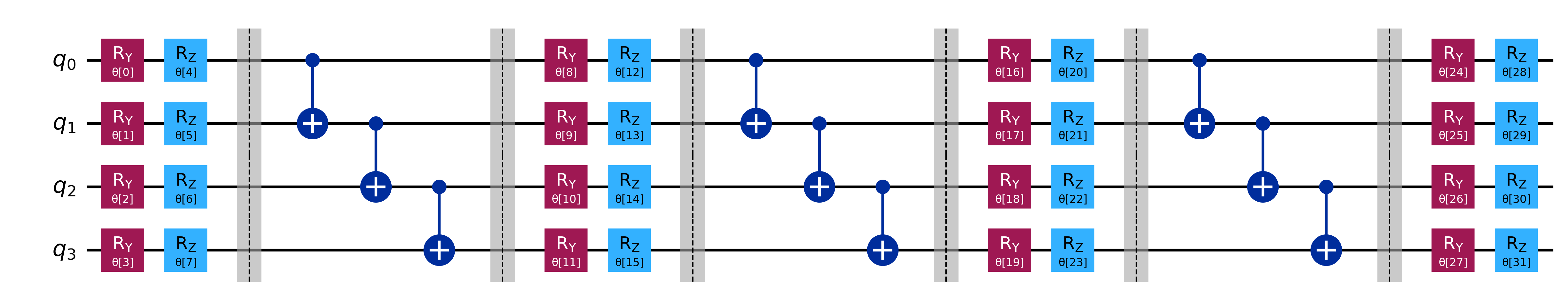}
    \caption{Ansatz after updating the VQE setup.}
    \label{fig:ansatz_best}
\end{figure}

In addition to the changes made based upon the previous experiments, we also increased the overall expressivity of our ansatz. This was done by increasing the number of rotation and entanglement layers by one, which increased the dimension of the parameter space from $12$ to $32$~\footnotemark. Despite having previously shown no strong dependence between the number of rotation gates and the accuracy, this experiment was done when all components were the same as the initialisation; as a result, we decided that the potential benefits of increased expressivity~\footnotemark, when combined with other amendments, was worth a potential increase in cost. These changes produce the ansatz shown in~{Fig}~\ref{fig:ansatz_best}.

\footnotetext{Whilst 32 is bigger than the Hilbert space dimension, 16. Only the optimiser interacts with the parameter space -- not the quantum system. This issue of parametrised subspaces having larger dimension than their parent Hilbert space is unique to small systems. In general, the number of parameters scales polynomial whilst the Hilbert space scales exponentially, guaranteeing a smaller subspace.}

\footnotetext{Although the results of Appendix~\ref{Apdx:extra_results} seem to contradict this choice, such results where gathered using the initial VQE setup. We decided to increase expressivity as we believed this would be of benefit when combined with other optimisations. We are now sceptical that this is true due to the large number of parameters.}

\begin{table}[h]
    \centering
    \begin{tabular}{ccc@{\hspace{1cm}}ccc}
        \toprule
        Run & $E_{0}^{\mathrm{VQE}}$ & $N_{\mathrm{pauli}}$
        & Run & $E_{0}^{\mathrm{VQE}}$ & $N_{\mathrm{pauli}}$ \\
        \midrule
        1 & $-4.23$ & $2590$ & 6  & $-3.97$ & $2500$ \\
        2 & $-3.67$ & $2790$ & 7  & $-4.21$ & $2800$ \\
        3 & $-4.18$ & $2740$ & 8  & $-3.77$ & $2570$ \\
        4 & $-2.95$ & $2660$ & 9  & $-4.02$ & $2680$ \\
        5 & $-3.90$ & $2360$ & 10 & $-4.33$ & $2340$ \\
        \bottomrule
    \end{tabular}
    \caption{Final VQE energy estimates and corresponding number of Pauli string evaluations over ten runs.}
    \label{tab:final_vqe_results}
\end{table}

\begin{figure}[h]
    \centering
    \includegraphics[width=1.0\linewidth]{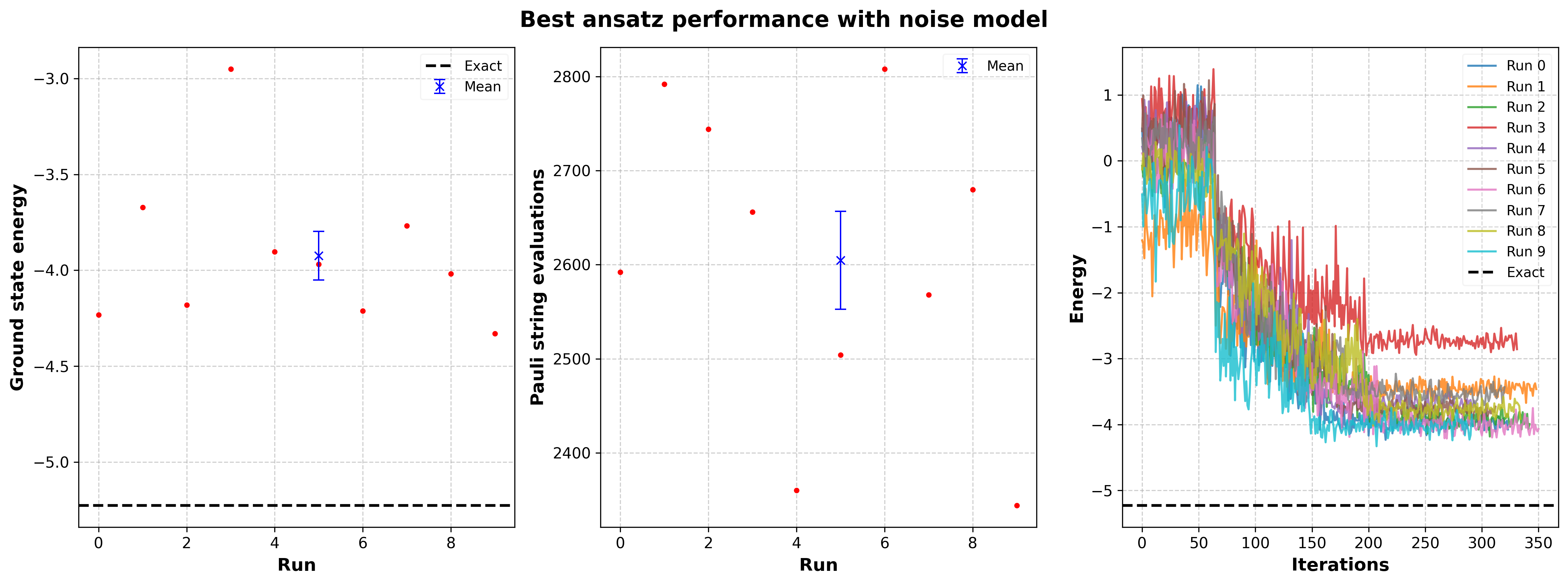}
    \caption{Results of the simulation using the optimised VQE setup.}
    \label{fig:best_VQE_results}
\end{figure}

From the ten independent VQE runs, the mean ground state energy is given as
\begin{equation}
    \label{eq:best_E0}
    \overline{E_{0}^{VQE}} = -3.92 \pm 0.13.
\end{equation}
We can now calculate its relative error:
\begin{equation}
\epsilon_{\mathrm{rel}} = \frac{|\overline{E_{0}^{\mathrm{VQE}}} - E_{0}^{\mathrm{exact}}|}
    {|E_{0}^{\mathrm{exact}}|} \times 100 = 
    \frac{|-3.92 - (-5.23)|}{|-5.23|} \times 100
    \approx 25\%.
\end{equation}
This relative error has roughly halved compared to the initial VQE setup, indicating that the changes we made have led to a more accurate result. This has been achieved via simple adjustments to the key VQE components, without utilising out more advanced techniques such as adaptive ansatz or error mitigation.~\footnote{No error mitigation could be performed on these results as there is no built in Qiskit methods for error mitigation when using a local simulator with the \texttt{estimator} class.}

The mean number of Pauli string evaluations has increased to
\begin{equation}
    \overline{N}_{\mathrm{Pauli}} = 2600 \pm 52 .
\end{equation}
This result corresponds to an increase by approximately a factor of five compared to the initial configuration. The reflects the increased computational cost required for a more expressive ansatz, further highlighting the trade-off between expressibility and cost, particular on NISQ devices where circuit executions are both time and noise limited.

\section{Demonstrating a quantum phase transition}\label{sec:VQE_QPT}

Utilising the optimised VQE setup, we now begin the simulation demonstrating a quantum phase transition. Just as we did in Sec.~\ref{sec:ED_PT}, we fix $J=1$ while sweeping $h$ from $0 \to 4$. For each $h$ value, we perform ten independent VQE runs with $1000$ shots each, using the \texttt{FakeManilaV2} noise model. In addition to the VQE setup from~{Sec.}~\ref{sec:best_VQE}, we also implement a specific type of measurement error mitigation, matrix-free error mitigation, otherwise labelled M3~\cite{PRXQuantum.2.040326}. M3 mitigation is a measurement mitigation technique that solves for corrected measurement probabilities using a dimensionality reduction step followed by either direct LU factorization or a preconditioned iterative method~\cite{PRXQuantum.2.040326}. This mitigation technique is particularly useful here as its associated Python library, \texttt{mthree}, can be applied automatically within VQE pipeline, even when using a local simulator; this contrasts against many other mitigation libraries, which cannot be readily integrated into local simulation workflows.

\begin{figure}[H]
    \centering
    \includegraphics[width=1.0\linewidth]{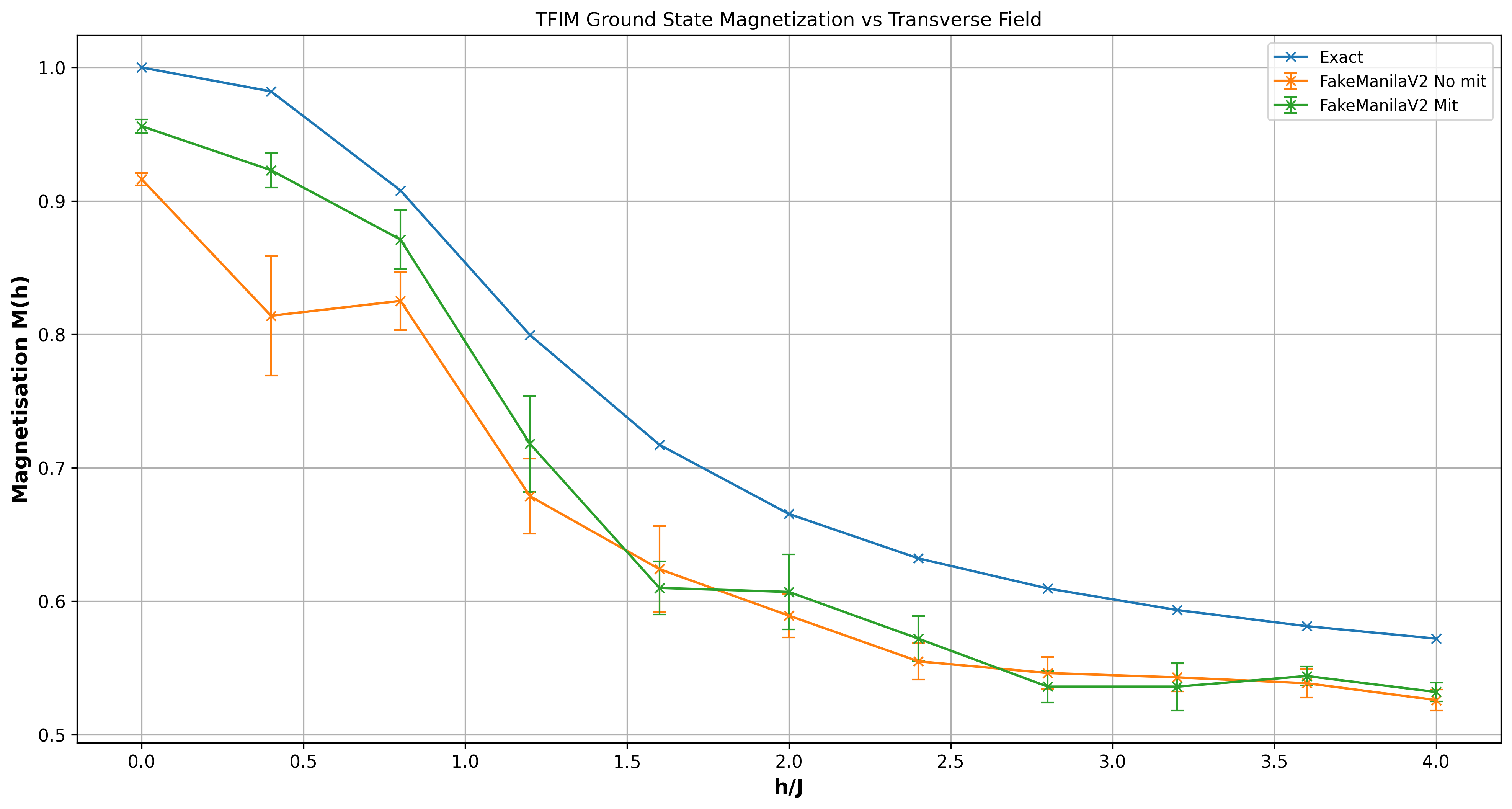}
    \caption{Results of the VQE both with and without error mitigation applied and compared to the result found using classical exact diagonalisation.}
    \label{fig:VQE_QPT}
\end{figure}

Firstly, we observe both results largely match the overall shape of the exact result. In addition, we see that M3 mitigation improves the results, particularly in the ordered phase, bringing us closer to the exact curve. This demonstrates the potential upsides of pairing error mitigation techniques with an algorithm being run on NISQ hardware. The noticeable improvements from M3 in the ordered ferromagnetic phase are likely due to the smaller quantum fluctuations. As a result, the measured states will contain fewer up-down superpositions, making the measurement process much easier even in the presence of quantum noise.

Interestingly, even when $h=0$, i.e no quantum noise, neither of the VQE curves are able to completely match that of the exact result. In this instance, the state contains only fully aligned spins, with no superpositions between spin configurations. Therefore, we would expect all results to agree. We conclude that the presence of noise is causing the offsets for both of the VQE results. This is an important observation in terms of its implications for quantum simulation. Even when the physical system can be well understood and predicted, on NISQ devices we should always expect some error in results due to noise, even when error mitigation is applied. Despite this, results like these still serve as proofs of concept that quantum systems can be successfully modelled on NISQ devices, providing a stepping stone on which future work can build. 

\section{Running on real hardware}\label{sec:hardware}
The final stage of this work was to run the VQE simulation on real quantum hardware to estimate the ground state energies, we return to the case where $h=1$ and $J=1$. As already mentioned, we use \texttt{ibmq\_fez} (version 1.3.37) which is a $156$ qubit device. IBM currently allows users a maximum of $10$ minutes of QPU time per $28$ day rolling cycle; to ensure we got results, we had to make some changes to the VQE initialisation. 

Whilst we kept the same VQE setup from before in terms of the ansatz, we had to change elements of the running process. Firstly, we reduced the number of shots to $500$ per iteration from $1000$. Additionally, we only allowed the optimiser a single iteration instead of $750$. Although this was the most substantial modification, it was necessary to obtain results within the ten-minute QPU usage allowance. To improve the reliability of future results, the number of optimisation iterations should be increased where QPU time permits. As this device contains $156$ physical qubits, we must transpile our current four-qubit ansatz to fit the device's qubit layout (see~{Fig.}~\ref{fig:fez_manila_qubit_layout}). One part of this  transpilation process is to convert the ansatz gates to the device's basis gates. Specifically, \texttt{ibmq\_fez} (version 1.3.37) has basis rotation gates $\{R_x, R_z, R_{zz}\}$ and basis entanglement gates Controlled$-Z$ ($CZ$)~\cite{IBM_fez}. After we transpile the new circuit contains: $60 \; R_Z$ gates, $6 \; CZ$ gates and $36$ SWAP gates. Clearly, this transpiled circuit is very different from our original ansatz and we shall see the effect this has on our results.

\begin{figure}[h]
    \centering
    \includegraphics[width=1.0\linewidth]{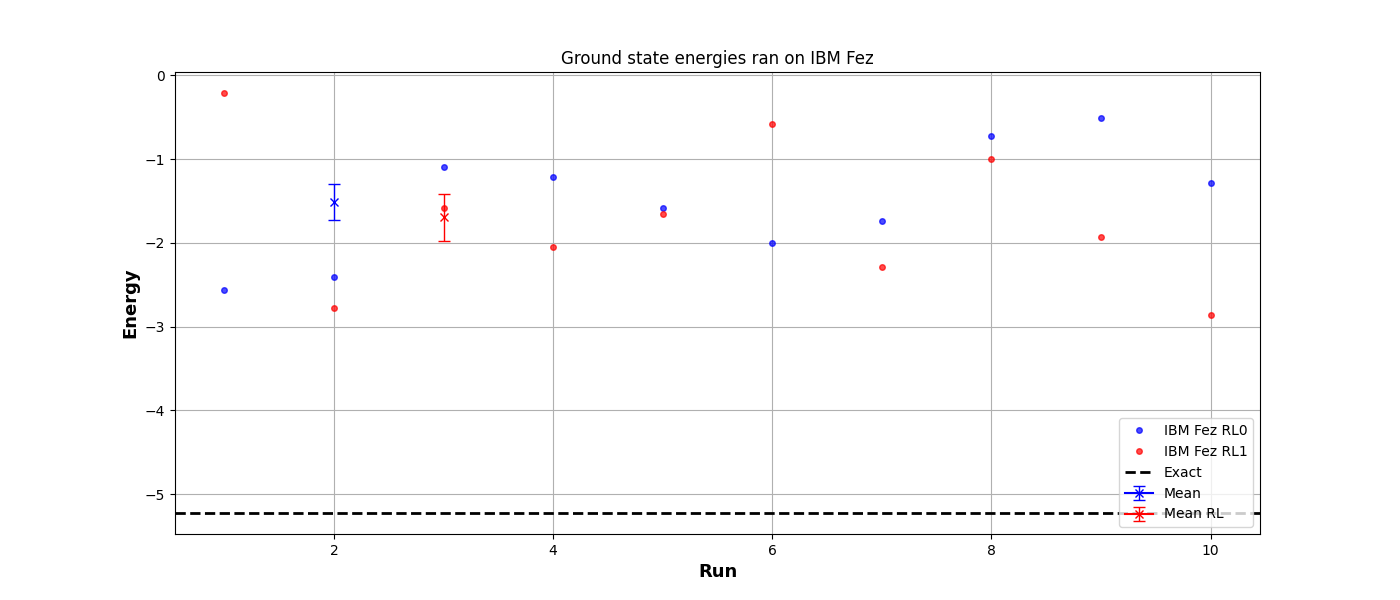}
    \caption{Ground state energy estimates ran on real hardware.}
    \label{fig:hardware_engy}
\end{figure}

Due to no longer using a local simulator, we are able to test the effects of error mitigation. We explore error mitigation using Qiskit's built-in resilience levels (RL)~\cite{Qiskit_RL}. Resilience levels are built in to Qiskit's \texttt{EstimatorV2} class and when used on real hardware allow the easy application of error mitigation. The resilience level can be easily set by setting the \texttt{resilience\_level} option a given number: $0$, $1$ or $2$. Each level  represents an increased amount of applied mitigation, with higher levels generating more accurate results, at the expense of longer processing times~\cite{Qiskit_RL}. Due to the given QPU time constraints, we test resilience level $0$ which uses no error mitigation techniques, and resilience level $1$ which provides Twirled Readout Error eXtinction (TREX) measurement twirling~{(see Ref.}~\cite{Cai_2023}~{for more details)}. We present our results in~{table}~\ref{tab:hardware_RL_comparison} with the plotted values in~{Fig.}~\ref{fig:hardware_engy}.

\begin{table}[h]
    \centering
    \begin{tabular}{cc c cc}
        \toprule
        \multicolumn{2}{c}{$RL=0$} & \hspace{1cm} & \multicolumn{2}{c}{$RL=1$} \\
        \midrule
        Run & Energy & & Run & Energy \\
        \midrule
        1  & $-2.56$ & & 1  & $-0.210$ \\
        2  & $-2.40$ & & 2  & $-2.78$ \\
        3  & $-1.09$ & & 3  & $-1.58$ \\
        4  & $-1.22$ & & 4  & $-2.05$ \\
        5  & $-1.58$ & & 5  & $-1.66$ \\
        6  & $-2.01$ & & 6  & $-0.582$ \\
        7  & $-1.74$ & & 7  & $-2.28$ \\
        8  & $-0.730$ & & 8  & $-0.998$ \\
        9  & $-0.512$ & & 9  & $-1.93$ \\
        10 & $-1.29$ & & 10 & $-2.86$ \\
        \midrule
        \multicolumn{2}{c}{$\overline{E_{RL=0}} = -1.51 \pm 0.21$}
        & &
        \multicolumn{2}{c}{$\overline{E_{RL=1}} = -1.69 \pm 0.28$} \\
        \multicolumn{2}{c}{$\epsilon_{\mathrm{rel}} = 0.71 \; (71\%)$}
        & &
        \multicolumn{2}{c}{$\epsilon_{\mathrm{rel}} = 0.68 \; (68\%)$} \\
        \bottomrule
    \end{tabular}
    \caption{Ground-state energy estimates from IBM Fez hardware runs for $RL=0$ and $RL=1$. The mean values are quoted with the standard error of the mean, and relative errors are calculated with respect to $E_{0}^{\mathrm{exact}}=-5.226$.}
    \label{tab:hardware_RL_comparison}
\end{table}

We observe only a slight improvement in results when going from $\text{RL}=0$ to $\text{RL}=1$. We believe this indicates that our main source of error is not due to readout error; rather, we suspect that the main sources of error are coming from the gates and/or classical optimisation process. If we were to increase the resilience level to $\text{RL}=2$, which would include zero noise extrapolation (ZNE)~\cite{Giurgica_Tiron_2020} and gate twirling~\cite{Cai_2023} in addition to TREX, we would expect a noticeable improvement in the accuracy of the ground state estimates. A similar effect was observed in~{Ref.}~\cite{Kerfoot:2026ngf} when moving from $\text{RL}=1$ to $\text{RL}=2$.

Additionally, we note that both real hardware results are our least accurate. This is to be expected, as in addition to the hardware and measurement error that has been modelled using a fake backend, the real hardware also has to evaluate a much larger ansatz, due to the transpilation process, for each expectation value. Although transpilation doesn't increase the number of randomised parameters, the number of gates does increase leading to a higher number of gate evaluation operations. As a result, we believe this will have an overall negative impact on the VQE efficiency due to the increased evaluation time. Despite this, results like these demonstrate that the TFIM can be simulated via quantum simulation and serve as a stepping stone for future implementations to improve upon.

\chapter{Conclusions}
\label{sec:conclusion}
\section{Conclusions and Outlook}
We have simulated one of the most famous models in condensed matter physics, the transverse field Ising model (TFIM), using a variational quantum algorithm, specifically the variational quantum eigensolver (VQE). The TFIM is a $1$D lattice model that evolves in time used for the study of magnetism. Famously, the TFIM exhibits a quantum phase transition (QPT) at zero temperature in the infinite lattice limit at some critical point.

We initially simulated the TFIM classically for a small number of spins, $N=2, 4$ and $8$, in these instances we was able to demonstrate the behaviour of the TFIM in its two known phases: ordered ferromagnetic and disordered paramagnetic. Additionally, we were able to demonstrate the trend towards a discontinuous QPT as the system size $N$ increased.

Following on from this, we implemented an initial VQE configuration and made an estimate of the ground state energy using a local noise simulator, achieving a relative error of $\sim 40\%$ compared to classical exact diagonalisation. We performed several different tests to optimise our VQE configuration. After making adjustments based on such tests, we were able to reduce the relative error to $\sim 25\%$. Due to software and time constraints, we were unable to pair these improvements with error mitigation techniques.

Using the optimised VQE setup we performed two additional tests. Firstly, still using a local simulator, we reproduced the QPT plot. We found that the overall shape and trend was present; however, classical exact diagonalisation was able to exhibit better physical behaviour. Implementing built-in readout error mitigation techniques did improve the results, particularly in the ordered ferromagnetic phase.

The last part of this work was to run our VQE configuration on real quantum hardware. For this, we focused on the ground state energy estimation, due to the IBM $10$ minute QPU usage limit. After running, we achieved a relative error of around $ 70\%$ with and without error mitigation, this is a significant error and far from desirable for future application. However, we believe these results can be significantly improved given the use of additional error mitigation techniques. In particular, we recommend implementing error mitigation techniques for gate errors such as gate twirling. Given more QPU time these error mitigation techniques could be ran in conjunction with the QPT plot. We believe doing this would lead to more accurately results and give better predictions of the physical behaviour, particularity near the critical point.

Additionally, we would like to explore further variational quantum algorithms for the simulation of the TFIM. Doing so would allow us to establish the most effective algorithm for this specific problem, which can then be scaled to larger systems. Specific algorithms we would like to test in the future include: quantum approximation optimisation algorithm (QAOA)~\cite{farhi2014quantum} and sample based Krylov quantum diagonalisation (SKQD) algorithm~\cite{yu2025quantum}.

While these results serve as a proof of concept, demonstrating that small quantum systems can be simulated using quantum computers, they do not outperform classical simulation methods. Achieving quantum advantage for the simulation of realistic quantum systems would require the treatment of much larger system sizes than those considered in this work, while also doing so faster or more accurately than the best available classical methods. In general, such simulations are expected to remain challenging on NISQ-era devices due to noise, limited qubit counts, and circuit-depth restrictions. It is therefore likely that large-scale, practically useful quantum simulations will require substantial improvements in hardware and error mitigation, and may ultimately depend on the development of fully fault-tolerant quantum computers.

\clearpage
\appendix 
\pagenumbering{roman}

\chapter{VQE optimisation results}

\label{Apdx:extra_results}
Below we present some summarised data for the extra experiments not included in the main text. In addition to this we present the full data for all experiments discussed within the main text.

\section{Varying ansatz gate repetitions}

\begin{figure}[h]
    \centering
    \includegraphics[width=1.0\linewidth]{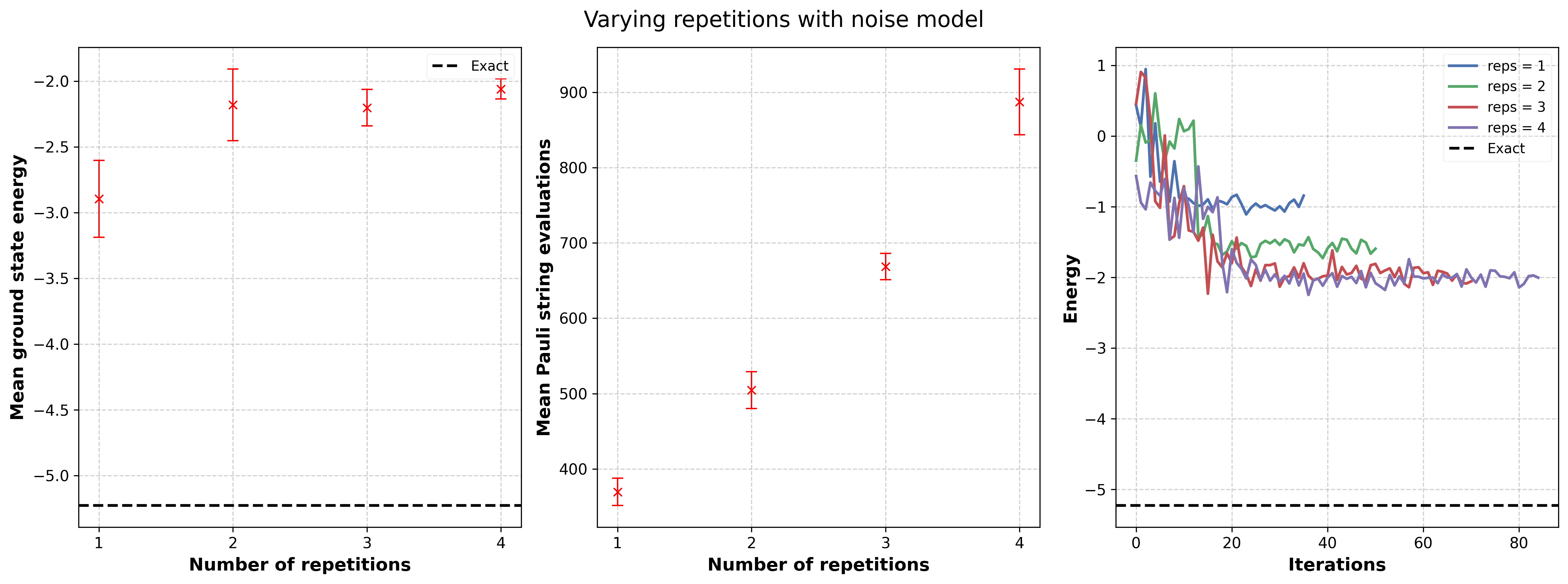}
    \caption{Results for varying the ansatz gates repetitions whilst keeping all other components fixed.}
    \label{fig:vary_rep}
\end{figure}

\begin{table}[H]
    \centering
    \label{tab:noisy_reps_results}
    \begin{tabular}{ccccc}
        \toprule
        Repetitions & $\overline{E_{0}^{VQE}}$ & $\Delta\overline{E_{0}^{VQE}}$ & $\overline{N_{pauli}}$ & $\Delta \overline{N_{pauli}}$ \\
        \midrule
        1 & $-2.895$ & $0.292$ & $369.6$ & $18.0$ \\
        2 & $-2.181$ & $0.272$ & $504.8$ & $24.4$ \\
        3 & $-2.201$ & $0.140$ & $668.8$ & $17.5$ \\
        4 & $-2.059$ & $0.076$ & $887.2$ & $43.5$ \\
        \bottomrule
    \end{tabular}
    \caption{Summary of noisy VQE results for different ansatz gate repetitions. Energy and Pauli evaluation uncertainties are reported as standard errors of the mean.}
\end{table}

\section{Varying shots}

\begin{figure}[H]
    \centering
    \includegraphics[width=1.0\linewidth]{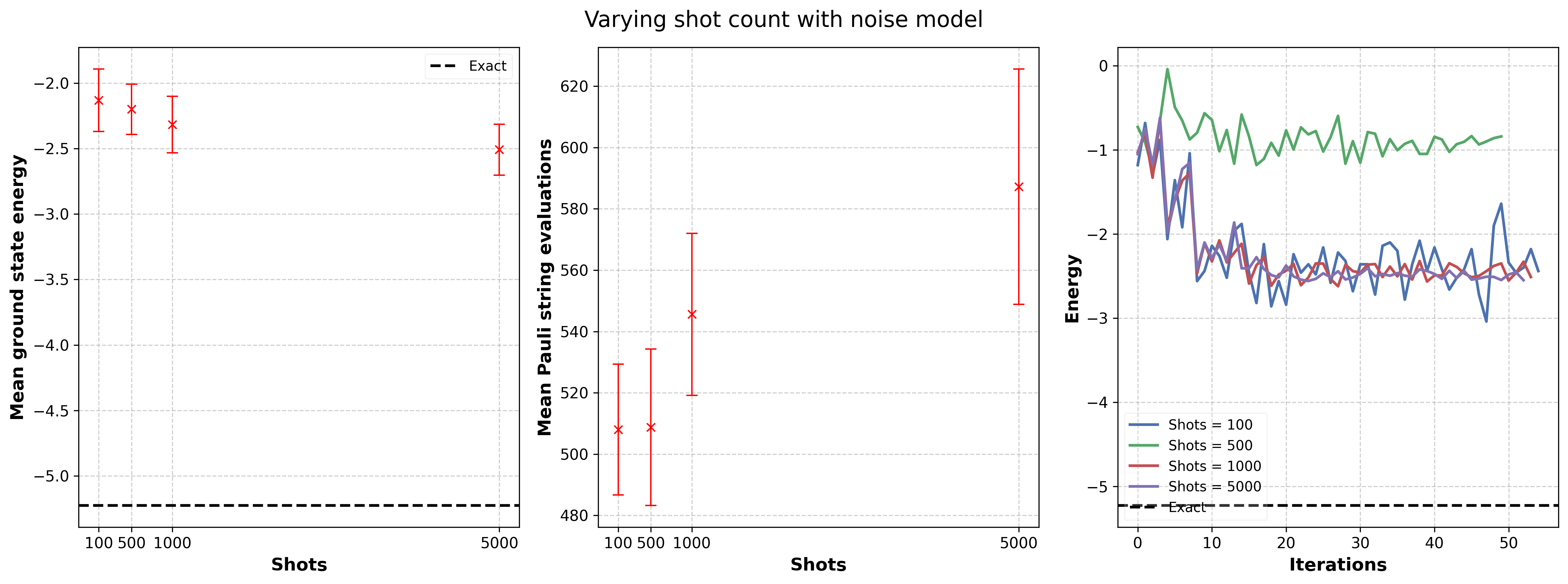}
    \caption{Results for varying the number of shots whilst keeping all other components fixed.}
    \label{fig:vary_shots}
\end{figure}

\begin{table}[H]
    \centering
    \label{tab:noisy_shots_results}
    \begin{tabular}{ccccc}
        \toprule
        Shots & $\overline{E_{0}^{VQE}}$ & $\Delta\overline{E_{0}^{VQE}}$ & $\overline{N_{pauli}}$ & $\Delta \overline{N_{pauli}}$ \\
        \midrule
        100  & $-2.130$ & $0.238$ & $508.0$ & $21.3$ \\
        500  & $-2.199$ & $0.192$ & $508.8$ & $25.5$ \\
        1000 & $-2.316$ & $0.215$ & $545.6$ & $26.4$ \\
        5000 & $-2.508$ & $0.194$ & $587.2$ & $38.4$ \\
        \bottomrule
    \end{tabular}
    \caption{Summary of noisy VQE results for different shot counts. Energy and Pauli evaluation uncertainties are reported as standard errors of the mean.}
\end{table}

\section{Full entanglements data}

\begin{longtable}{cccc}
    \label{tab:appendix_full_entanglement_results} \\
    \toprule
    Entanglement strategy & Run & $E_{0}^{\mathrm{VQE}}$ & $N_{\mathrm{pauli}}$ \\
    \midrule
    \endfirsthead

    \toprule
    Entanglement strategy & Run & $E_{0}^{\mathrm{VQE}}$ & $N_{\mathrm{pauli}}$ \\
    \midrule
    \endhead

    \midrule
    \multicolumn{4}{r}{Continued on next page} \\
    \endfoot

    \endlastfoot

    Circular & 1  & $-3.263$ & $568$ \\
    Circular & 2  & $-2.354$ & $608$ \\
    Circular & 3  & $-1.449$ & $536$ \\
    Circular & 4  & $-2.386$ & $632$ \\
    Circular & 5  & $-2.585$ & $432$ \\
    Circular & 6  & $-3.247$ & $536$ \\
    Circular & 7  & $-3.485$ & $648$ \\
    Circular & 8  & $-2.873$ & $528$ \\
    Circular & 9  & $-2.753$ & $528$ \\
    Circular & 10 & $-1.343$ & $440$ \\
    \midrule
    \multicolumn{2}{c}{Mean circular} & $-2.574 \pm 0.230$ & $545.6 \pm 22.9$ \\
    \midrule

    Linear & 1  & $-4.142$ & $568$ \\
    Linear & 2  & $-3.654$ & $592$ \\
    Linear & 3  & $-3.085$ & $528$ \\
    Linear & 4  & $-3.646$ & $432$ \\
    Linear & 5  & $-4.074$ & $456$ \\
    Linear & 6  & $-3.177$ & $616$ \\
    Linear & 7  & $-4.114$ & $552$ \\
    Linear & 8  & $-4.280$ & $560$ \\
    Linear & 9  & $-3.576$ & $488$ \\
    Linear & 10 & $-2.474$ & $432$ \\
    \midrule
    \multicolumn{2}{c}{Mean linear} & $-3.622 \pm 0.181$ & $522.4 \pm 21.1$ \\
    \midrule

    Reverse linear & 1  & $-4.124$ & $600$ \\
    Reverse linear & 2  & $-3.838$ & $576$ \\
    Reverse linear & 3  & $-3.680$ & $712$ \\
    Reverse linear & 4  & $-2.264$ & $496$ \\
    Reverse linear & 5  & $-3.293$ & $496$ \\
    Reverse linear & 6  & $-3.698$ & $528$ \\
    Reverse linear & 7  & $-3.904$ & $504$ \\
    Reverse linear & 8  & $-4.088$ & $632$ \\
    Reverse linear & 9  & $-2.525$ & $440$ \\
    Reverse linear & 10 & $-3.680$ & $416$ \\
    \midrule
    \multicolumn{2}{c}{Mean reverse linear} & $-3.509 \pm 0.201$ & $540.0 \pm 28.6$ \\
    \midrule

    Pairwise & 1  & $-3.918$ & $456$ \\
    Pairwise & 2  & $-3.453$ & $536$ \\
    Pairwise & 3  & $-3.491$ & $504$ \\
    Pairwise & 4  & $-3.748$ & $592$ \\
    Pairwise & 5  & $-3.618$ & $440$ \\
    Pairwise & 6  & $-2.657$ & $528$ \\
    Pairwise & 7  & $-4.096$ & $528$ \\
    Pairwise & 8  & $-3.550$ & $544$ \\
    Pairwise & 9  & $-3.976$ & $472$ \\
    Pairwise & 10 & $-4.152$ & $760$ \\
    \midrule
    \multicolumn{2}{c}{Mean pairwise} & $-3.666 \pm 0.137$ & $536.0 \pm 28.7$ \\
    \midrule

    SCA & 1  & $-3.131$ & $592$ \\
    SCA & 2  & $-3.542$ & $512$ \\
    SCA & 3  & $-3.463$ & $576$ \\
    SCA & 4  & $-3.502$ & $768$ \\
    SCA & 5  & $-1.940$ & $512$ \\
    SCA & 6  & $-1.081$ & $416$ \\
    SCA & 7  & $-2.509$ & $648$ \\
    SCA & 8  & $-3.205$ & $584$ \\
    SCA & 9  & $-3.119$ & $608$ \\
    SCA & 10 & $-3.802$ & $512$ \\
    \midrule
    \multicolumn{2}{c}{Mean SCA} & $-2.929 \pm 0.268$ & $572.8 \pm 30.0$ \\
    \bottomrule
    \caption{Summary of full VQE results for different entanglement types. Energy and Pauli evaluation uncertainties are reported as standard errors of the mean.}
\end{longtable}

\section{Full optimisers data}
\begin{longtable}{cccc}
    \toprule
    Optimiser & Run & $E_{0}^{\mathrm{VQE}}$ & $N_{\mathrm{pauli}}$ \\
    \midrule
    \endfirsthead

    \toprule
    Optimiser & Run & $E_{0}^{\mathrm{VQE}}$ & $N_{\mathrm{pauli}}$ \\
    \midrule
    \endhead

    \midrule
    \multicolumn{4}{r}{Continued on next page} \\
    \endfoot

    \endlastfoot

    COBYLA & 1  & $-2.695$ & $648$ \\
    COBYLA & 2  & $-1.576$ & $544$ \\
    COBYLA & 3  & $-1.431$ & $472$ \\
    COBYLA & 4  & $-1.590$ & $408$ \\
    COBYLA & 5  & $-1.864$ & $456$ \\
    COBYLA & 6  & $-3.441$ & $552$ \\
    COBYLA & 7  & $-2.156$ & $488$ \\
    COBYLA & 8  & $-2.086$ & $552$ \\
    COBYLA & 9  & $-1.816$ & $464$ \\
    COBYLA & 10 & $-1.361$ & $472$ \\
    \midrule
    \multicolumn{2}{c}{Mean COBYLA} & $-2.002 \pm 0.203$ & $505.6 \pm 21.7$ \\
    \midrule

    COBYQA & 1  & $-3.570$ & $1120$ \\
    COBYQA & 2  & $-3.680$ & $1008$ \\
    COBYQA & 3  & $-2.893$ & $1008$ \\
    COBYQA & 4  & $-3.265$ & $1016$ \\
    COBYQA & 5  & $-3.407$ & $912$ \\
    COBYQA & 6  & $-3.277$ & $1032$ \\
    COBYQA & 7  & $-3.249$ & $888$ \\
    COBYQA & 8  & $-3.379$ & $904$ \\
    COBYQA & 9  & $-3.566$ & $960$ \\
    COBYQA & 10 & $-3.293$ & $1144$ \\
    \midrule
    \multicolumn{2}{c}{Mean COBYQA} & $-3.358 \pm 0.070$ & $999.2 \pm 27.4$ \\
    \midrule

    BFGS & 1  & $1.592$  & $3144$ \\
    BFGS & 2  & $0.853$  & $2280$ \\
    BFGS & 3  & $0.951$  & $2592$ \\
    BFGS & 4  & $-0.332$ & $2488$ \\
    BFGS & 5  & $-1.259$ & $2376$ \\
    BFGS & 6  & $-0.346$ & $3392$ \\
    BFGS & 7  & $-0.671$ & $1856$ \\
    BFGS & 8  & $-0.667$ & $2064$ \\
    BFGS & 9  & $-0.711$ & $3008$ \\
    BFGS & 10 & $-0.953$ & $2744$ \\
    \midrule
    \multicolumn{2}{c}{Mean BFGS} & $-0.154 \pm 0.299$ & $2594.4 \pm 153.3$ \\
    \midrule

    L-BFGS-B & 1  & $1.748$  & $2184$ \\
    L-BFGS-B & 2  & $1.039$  & $3016$ \\
    L-BFGS-B & 3  & $1.195$  & $1768$ \\
    L-BFGS-B & 4  & $-0.342$ & $2600$ \\
    L-BFGS-B & 5  & $-1.079$ & $2600$ \\
    L-BFGS-B & 6  & $-0.054$ & $3120$ \\
    L-BFGS-B & 7  & $-0.408$ & $2808$ \\
    L-BFGS-B & 8  & $-0.569$ & $3120$ \\
    L-BFGS-B & 9  & $-0.713$ & $2496$ \\
    L-BFGS-B & 10 & $-0.817$ & $1872$ \\
    \midrule
    \multicolumn{2}{c}{Mean L-BFGS-B} & $-0.000 \pm 0.308$ & $2558.4 \pm 154.4$ \\

    \bottomrule
    \caption{Summary of full noisy VQE results for different classical optimisers. Energy and Pauli evaluation uncertainties are reported as standard errors of the mean.}
    \label{tab:appendix_full_optimizer_results}
\end{longtable}

\section{Full rotation gates data}

\begin{longtable}{cccc}
    \toprule
    Rotation block & Run & $E_{0}^{\mathrm{VQE}}$ & $N_{\mathrm{pauli}}$ \\
    \midrule
    \endfirsthead

    \toprule
    Rotation block & Run & $E_{0}^{\mathrm{VQE}}$ & $N_{\mathrm{pauli}}$ \\
    \midrule
    \endhead

    \midrule
    \multicolumn{4}{r}{Continued on next page} \\
    \endfoot

    \endlastfoot

    $R_y$ & 1  & $-2.238$ & $568$ \\
    $R_y$ & 2  & $-2.282$ & $648$ \\
    $R_y$ & 3  & $-1.534$ & $544$ \\
    $R_y$ & 4  & $-2.190$ & $528$ \\
    $R_y$ & 5  & $-2.621$ & $408$ \\
    $R_y$ & 6  & $-3.107$ & $520$ \\
    $R_y$ & 7  & $-3.173$ & $696$ \\
    $R_y$ & 8  & $-1.930$ & $584$ \\
    $R_y$ & 9  & $-2.791$ & $488$ \\
    $R_y$ & 10 & $-1.195$ & $416$ \\
    \midrule
    \multicolumn{2}{c}{Mean $R_y$} & $-2.306 \pm 0.203$ & $540.0 \pm 28.8$ \\
    \midrule

    $R_y, R_x$ & 1  & $-2.232$ & $1064$ \\
    $R_y, R_x$ & 2  & $-2.609$ & $1152$ \\
    $R_y, R_x$ & 3  & $-2.641$ & $1024$ \\
    $R_y, R_x$ & 4  & $-3.119$ & $1648$ \\
    $R_y, R_x$ & 5  & $-2.629$ & $1040$ \\
    $R_y, R_x$ & 6  & $-2.681$ & $1024$ \\
    $R_y, R_x$ & 7  & $-2.761$ & $1152$ \\
    $R_y, R_x$ & 8  & $-2.370$ & $992$ \\
    $R_y, R_x$ & 9  & $-1.762$ & $896$ \\
    $R_y, R_x$ & 10 & $-1.866$ & $960$ \\
    \midrule
    \multicolumn{2}{c}{Mean $R_y, R_x$} & $-2.467 \pm 0.132$ & $1095.2 \pm 66.2$ \\
    \midrule

    $R_y, R_z$ & 1  & $-1.233$ & $856$ \\
    $R_y, R_z$ & 2  & $-2.735$ & $1104$ \\
    $R_y, R_z$ & 3  & $-2.446$ & $1056$ \\
    $R_y, R_z$ & 4  & $-2.208$ & $1064$ \\
    $R_y, R_z$ & 5  & $-2.751$ & $1360$ \\
    $R_y, R_z$ & 6  & $-3.111$ & $1480$ \\
    $R_y, R_z$ & 7  & $-2.452$ & $1232$ \\
    $R_y, R_z$ & 8  & $-2.875$ & $968$ \\
    $R_y, R_z$ & 9  & $-3.213$ & $1208$ \\
    $R_y, R_z$ & 10 & $-1.654$ & $1152$ \\
    \midrule
    \multicolumn{2}{c}{Mean $R_y, R_z$} & $-2.468 \pm 0.198$ & $1148.0 \pm 57.8$ \\
    \midrule

    $R_z, R_x$ & 1  & $-1.910$ & $1344$ \\
    $R_z, R_x$ & 2  & $-1.964$ & $936$ \\
    $R_z, R_x$ & 3  & $-1.431$ & $1112$ \\
    $R_z, R_x$ & 4  & $-2.446$ & $1120$ \\
    $R_z, R_x$ & 5  & $-1.790$ & $1152$ \\
    $R_z, R_x$ & 6  & $-2.156$ & $1016$ \\
    $R_z, R_x$ & 7  & $-2.266$ & $1112$ \\
    $R_z, R_x$ & 8  & $-1.888$ & $1032$ \\
    $R_z, R_x$ & 9  & $-2.390$ & $1040$ \\
    $R_z, R_x$ & 10 & $-2.312$ & $1016$ \\
    \midrule
    \multicolumn{2}{c}{Mean $R_z, R_x$} & $-2.055 \pm 0.100$ & $1088.0 \pm 35.1$ \\
    \midrule

    $R_y, R_x, R_z$ & 1  & $-1.892$ & $1584$ \\
    $R_y, R_x, R_z$ & 2  & $-2.807$ & $1552$ \\
    $R_y, R_x, R_z$ & 3  & $-1.556$ & $1456$ \\
    $R_y, R_x, R_z$ & 4  & $-1.694$ & $1408$ \\
    $R_y, R_x, R_z$ & 5  & $-2.747$ & $1664$ \\
    $R_y, R_x, R_z$ & 6  & $-3.015$ & $1408$ \\
    $R_y, R_x, R_z$ & 7  & $-2.537$ & $1592$ \\
    $R_y, R_x, R_z$ & 8  & $-3.073$ & $1440$ \\
    $R_y, R_x, R_z$ & 9  & $-2.470$ & $1736$ \\
    $R_y, R_x, R_z$ & 10 & $-2.719$ & $1896$ \\
    \midrule
    \multicolumn{2}{c}{Mean $R_y, R_x, R_z$} & $-2.451 \pm 0.173$ & $1573.6 \pm 50.1$ \\

    \bottomrule
    \caption{Summary of full noisy VQE results for different ansatz rotation blocks. Energy and Pauli evaluation uncertainties are reported as standard errors of the mean.}
    \label{tab:appendix_full_rotation_results}
\end{longtable}

\bibliographystyle{unsrtnat}


\bibliography{References}
\end{document}